\documentclass[a4paper]{jpconf}

\usepackage{graphicx}				
\usepackage{xspace}				
\usepackage{colortbl}				
\usepackage{amssymb,amstext}			

\newcommand{\dmu}{\textsl{DMU}\xspace}
\newcommand{\esa}{\textsl{ESA}\xspace}
\newcommand{\ieec}{\textsl{IEEC}\xspace}
\newcommand{\lisa}{\textsl{LISA}\xspace}
\newcommand{\lpf}{\textsl{\mbox{LPF}}\xspace}
\newcommand{\ltp}{\textsl{LTP}\xspace}
\newcommand{\nasa}{\textsl{NASA}\xspace}

\newcommand{\menta}[1]{\cite{Lobo_#1}}

\newcommand{\bma}[1]{\mbox{\boldmath${#1}\/$}}
\newcommand{\bOx}{\raisebox{-1.5 pt}{\Large $\Box$}}

\begin{document}

\title{\textsl{LISA} and \textsl{LISA PathFinder}, the endeavour to detect
low frequency GWs}

\author{H Ara\'ujo$^1$, C Boatella$^2$, M Chmeissani$^3$, A Conchillo$^2$,
E Garc\'\i a-Berro$^{2,4}$, C Grimani$^5$, W Hajdas$^6$,
A Lobo$^{2,7,}$\footnote[10]{To whom correspondence should be addressed.},
Ll Mart\'\i nez$^8$, M Nofrarias$^2$, JA Ortega$^2$, C Puigdengoles$^3$,
J Ramos-Castro$^9$, J Sanju\'an$^2$, P Wass$^1$ and X Xirgu$^2$}

\address{$^1$ Blackett Laboratory, Imperial College London, Prince Consort
	Road, London SW7 2BW, UK}
\address{$^2$ Institut d'Estudis Espacials de Catalunya ({\sl IEEC\/}),
	Edifici {\sl Nexus}, Gran Capit\`a~2--4, 08034 Barcelona, Spain}
\address{$^3$ Institut de F\'\i sica d'Altes Energies ({\sl IFAE\/}),
	Edifici C, Universitat Aut\`onoma de Barcelona, 08193 Bellaterra,
	(Barcelona), Spain}
\address{$^4$ Departament de F\'\i sica Aplicada, Universitat Polit\`ecnica
	de Catalunya, Escola Polit\`ecnica Superior de Castelldefels,
	Avda.\ Canal Ol\'\i mpic s/n, 08860 Castelldefels, Spain}
\address{$^5$ Universit\`a degli Studi di Urbino, and {\sl INFN\/} Florence,
	Istituto di Fisica, Via Santa Chiara 27, 61029 Urbino, Italy}
\address{$^6$ Department of Particles and Matter, Paul Scherrer Institut,
	ODRA 120, 5232 Villigen, Switzerland}
\address{$^7$ Instituto de Ciencias del Espacio, {\sl CSIC}, Campus
	{\sl UAB}, Facultat de Ci\`encies, Torre C-5 Parell, 2$^a$ Planta,
	E-08193 Bellaterra (Cerdanyola del Vall\`es), Barcelona, Spain}
\address{$^8$ {\sl At\'\i pic}, Parc Tecnol\`ogic del Vall\`es,
	08290 Cerdanyola del Vall\`es, Barcelona, Spain}
\address{$^9$ Departament d'Enginyeria Electr\`onica, {\sl UPC},
	Campus Nord, Edif.\ C4, Jordi Girona 1--3, 08034 Barcelona, Spain}

\ead{lobo@ieec.fcr.es}

\begin{abstract}
This is a review about \lisa and its technology demonstrator, \lisa
{\sl PathFinder}. We first describe the conceptual problems which need
to be overcome in order to set up a working interferometric detector
of low frequency Gravitational Waves (GW), then summarise the solutions
to them as currently conceived by the \lisa mission team. This will
show that some of these solutions require new technological abilities
which are still under development, and which need proper test before
being fully implemented. \lisa {\sl PathFinder\/} (\lpf) is the the
testbed for such technologies. The final part of the paper will address
the ideas and concepts behind the {\sl PathFinder\/} as well as their
impact on \lisa.
\end{abstract}

\section{Introduction}

Gravitational Waves (GW), i.e., radiant gravitational fields, were
absent in Newton's classical theory of gravity. For over two centuries,
though, nobody was missing them. It was only with the advent of Einstein's
Special Theory of Relativity~\menta{dover} in the early twentieth century
that such absence began to create some distress. It was indeed very odd,
it appeared to physicists at that time, that gravitation should
\emph{instantly} propagate to even the remotest places in the Universe,
since this was blatantly against the principle of causality just born
out of the discovery of the limit speed, $c$ ---the speed of light in
empty space.

Let us recall Poisson's equation for the gravitational potential:
\begin{equation}
 \nabla\phi({\bf x},t) = 4\pi G\,\varrho({\bf x},t)
 \label{Lobo_eq.1}
\end{equation}
whose formal solution is
\begin{equation}
 \phi({\bf x},t) = -G\,\int\,\frac{\varrho({\bf x'},t)}{|{\bf x}-{\bf x'}|}\,
 d^3x'
 \label{Lobo_eq.2}
\end{equation}

We observe in this formula, relating the gravitational potential
$\phi({\bf x},t)$ to the density of gravitating matter $\varrho({\bf x},t)$,
that there are no differential coefficients involving the time variable~$t$,
which merely plays the role of a \emph{parameter} in the equation. This
means that any changes in the source density happening at time
$t\/$\,=\,$t_0$, say, are \emph{also} felt at time $t\/$\,=\,$t_0$ by
the gravitational potential, no matter \emph{where}. In other words,
the value of the gravitational potential $\phi$ varies in time exactly
in parallel with the time variations of the generating density $\varrho$,
and does so for \emph{all} values of the field space coordinate {\bf x}.
We thus see that Newton's theory predicts that gravitational perturbations,
hence gravitational energy, too, travel across empty space with
\emph{infinite speed}.

A very interesting account of the endeavours in pursuit of the solution
to the riddle to build a \emph{causal} theory of gravity can be found
e.g.\ in a renowned book by A Pais~\menta{pais}. Let us simply recall
here that Einstein's General theory of Relativity (GR), the usually
accepted final solution, predicts that gravitational waves do indeed
exist, which travel at the speed of light in otherwise flat empty space,
have \emph{two} polarisation degrees of freedom, and are \emph{transverse}
to the propagation direction. In a suitable quasi-Lorentzian coordinate
system, Einstein's vacuum GW equations read~\menta{wein}
\begin{equation}
 \bOx\,\left(h_{\mu\nu}-\frac{1}{2}\,\eta_{\mu\nu}\,
 \eta^{\rho\sigma}h_{\rho\sigma}\right) = 0
 \label{Lobo_eq.3}
\end{equation}
where $h_{\mu\nu}$ are small perturbations to the flat Lorentzian geometry,
$\eta_{\mu\nu}$:
\begin{equation}
 g_{\mu\nu}({\bf x},t) = \eta_{\mu\nu} + h_{\mu\nu}({\bf x},t)\ ,\quad
 |h_{\mu\nu}({\bf x},t)|\ll 1
 \label{Lobo_eq.4}
\end{equation}

The relationship between the GW amplitudes and their sources, i.e.,
the equivalent of equation~\eref{Lobo_eq.2}, is also textbook material
\menta{ll75}. In the far zone~\menta{kip}, it is given by the
\emph{quadrupole formula}
\begin{equation}
 h_{ij}({\bf x},t) = \frac{4G}{c^4|{\bf x}|}\;{\cal P}_{ijkl}\;
 \ddot{Q}_{kl}(t-|{\bf x}|/c)
 \label{Lobo_eq.5}
\end{equation}
where ${\cal P}_{ijkl}$\,=\,${\cal P}_{ik}{\cal P}_{jl}$\,$-$\,%
 $(1/2)\,{\cal P}_{ij}{\cal P}_{kl}$, with
${\cal P}_{ij}$\,=\,$\delta_{ij}$\,$-$\,$x_ix_j/|{\bf x}|^2$, is the
\emph{transverse-traceless} (TT) projection operator, and
\begin{equation}
 Q_{ij}(t)\equiv\int_{\rm Source}\,\left(
 x_ix_j - \frac{1}{3}\,|{\bf x}|^2\delta_{ij}\right)
 \varrho({\bf x},t)\,d^3x
 \label{Lobo_eq.6}
\end{equation}
is the source's quadrupole moment.

Equation~\eref{Lobo_eq.5} reflects the \emph{causal} nature of GWs in the
\emph{retarded time} argument in the rhs. For \emph{gravitationally bound}
systems, the formula can also be easily used to find an order of magnitude
estimate of the amplitude of the GW emission of a given source: for one of
mass $M\/$ and linear size $\ell$, which is observed from a distance $R\/$
away, it gives
\begin{equation}
 h\sim\frac{\ell}{R}\,\left(\frac{GM}{\ell c^2}\right)^{\!\!2}
 \label{Lobo_eq.7}
\end{equation}

Note that the term inside the brackets is the ratio between the source's
\emph{Schwarzschild radius} and its actual size, while the other term is
the ratio between the size of the source and its distance to the observer.
The combination of these two factors results in extremely small numbers
for GW amplitudes, at most $\sim$10$^{-18}$ in the most favourable
conditions of source intensity and likely distance. Any terrestrial, or
even Solar System sources conceivable generate much weaker signals. This
is most likely the reason why Celestial Mechanics managed to live a long
and successful life without GWs in its body of doctrine.

But once we know that GWs \emph{do} exist~\menta{taylor}, we discover they
can open a whole \emph{new window} for the observation of the Universe, so
far naturally unexplored, and with surely unsuspected capabilities in
store. The first experimental attempts to detect GWs with a dedicated
instrument are almost half a century old, and date back to the early
1960s, when J Weber designed, built and operated two very sensitive
cylindrical bars~\menta{weber}. 

\begin{figure}[t]
\centering
\includegraphics[width=14cm]{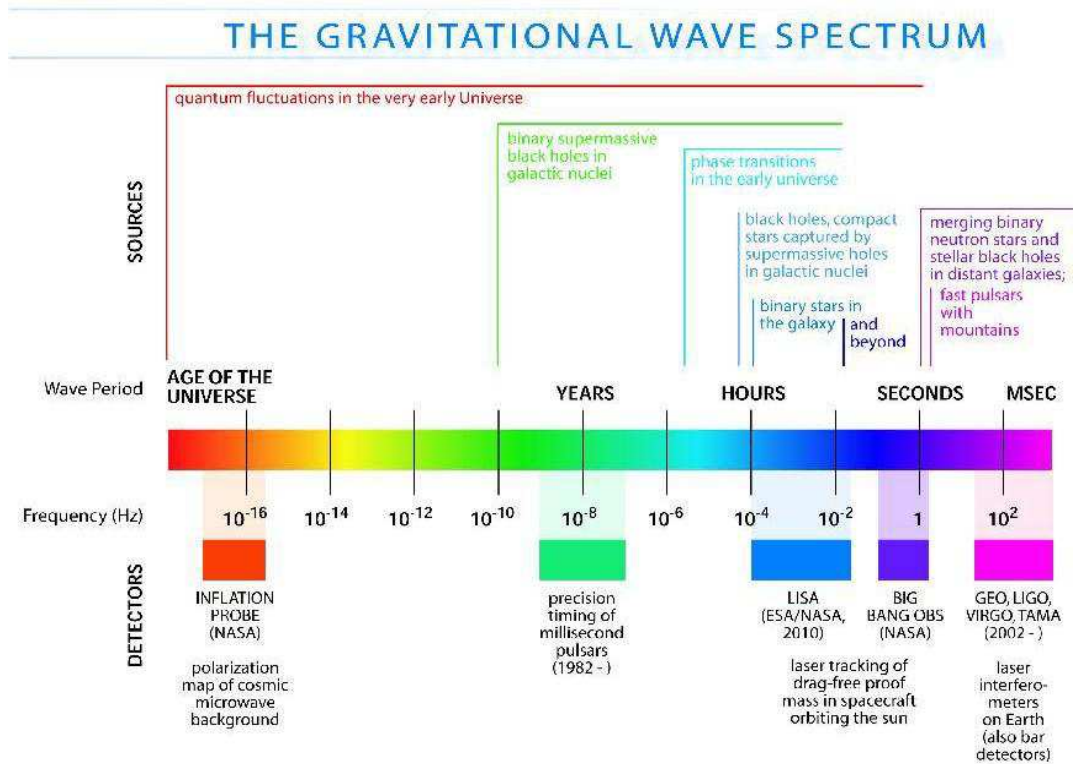}
\caption{The frequency spectrum of GWs.
\label{Lobo_fig.1}}
\end{figure}

Weber's claims of GW sightings eventually proved unconvincing, but they
did foster new research activity in the field of GW detectors. Current
earth based detectors are orders of magnitude more sensitive than
Weber's~\menta{naut,Lobo_aur}, and new concept~\menta{ron} interferometric
antennas have taken over Weber's idea of acoustic sensing in favour of
optical techniques. Such detectors as LIGO~\menta{ligo}, VIRGO~\menta{virgo}
or GEO-600~\menta{geo} are in very advanced states of development, and
should be shortly generating true GW observatory data.

Earth based detectors share however a common sensitivity limitation, set
by unavoidable seismic and gravity gradient noise~\menta{seism}. These
make basically unreachable the frequency band below $\sim$10\,Hz, with
the consequence that low frequency GWs cannot possibly be observed from
the earth's surface. If we wish to see such GWs then we have to move out
to space, far from terrestrial accidents. This is the reason for \lisa,
the first low frequency GW detector, which should fly in the time frame
of 2015 with good sensitivity in the band around 1~mHz~\menta{lisaweb}.

There is a qualitative argument that low frequency GW signals should be
more abundant than high frequency ones ---where ``high frequency'' means
around and above~1~kHz. This is that large scale mass motions have
typically long time scales, too, mostly far from the fraction-of-a-second
scale. Figure~\ref{Lobo_fig.1} shows the frequency spectrum of GWs as we
currently conceive it, and roughly confirms that qualitative argument.
The figure also carries indications of the detectors sensitive in each
frequency region: as can be seen, \lisa clearly covers a wider range of
likely GW sources than cover any earth based antennas. This makes of \lisa
a specially attractive project, as it will provide the only means we
have to access such very interesting signals. In addition, \lisa's
potential for discovery of new unforeseen sources is also vigorously
there, and it will add to the potential of earth based instruments.

This paper is a review about \lisa workings, from the mission principles
and concept to its technology precursor \lisa PathFinder (\lpf). Its
organisation follows sequentially these topics.

\section{\textsl{LISA}}

\begin{figure}[b]
\centering
\includegraphics[width=9cm]{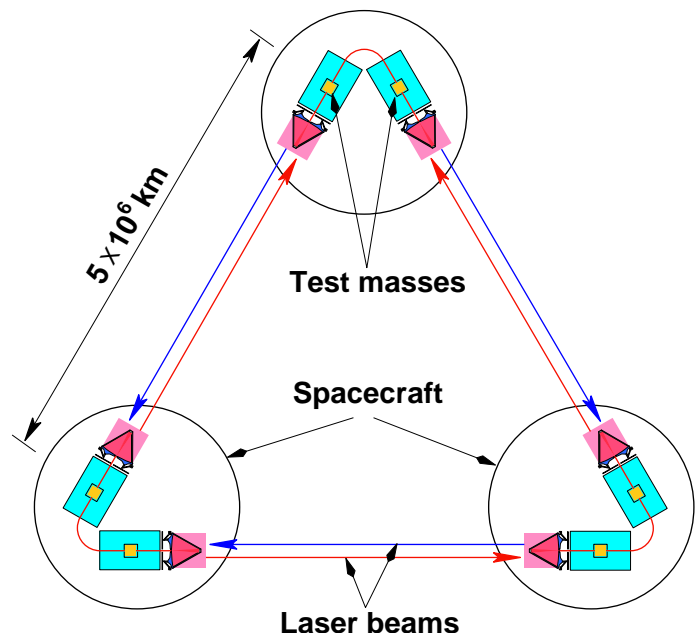}
\caption{\lisa's equilateral triangle configuration.
\label{Lobo_fig.2}}
\end{figure}

As is well known~\menta{cqg92}, interferometric detection of GWs
requires the interferometer arm-length to be close to $\lambda_{\rm GW}/2$.
If we aim at GWs of frequencies around 1~mHz, we correspondingly need
arm-lengths in the order of several million kilometres. For \lisa the
option taken is
\[
 \text{\textsl{LISA}'s arm-length} = 5\times 10^6\ {\rm kilometres}
\]

The configuration is shown in figure~\ref{Lobo_fig.2}: it is a constellation
of three spacecraft occupying the vertexes of an equilateral triangle.
There are a few important facts about this configuration which we now
summarise briefly.

\subsection{Active mirrors}

The 5 million km arm-length poses a problem to classical Michelson
interferometry, which is the following: even if the laser beam is
collimated to high precision, there is a minimum \emph{beam divergence}
which cannot be avoided. For a 1~watt laser, as foreseen for \lisa,
this limit is about 4$\times$10$^{-6}$ rad. The consequence of this
divergence is that, after 5 million km, the initial small spot of the
laser source has become a considerable spot of about 20 km in
diameter\ldots\ Therefore only a very small fraction of the emitted
light is actually collected by the remote mirror, which is a telescope
of only 40~cm in diameter. If light were simply reflected back to the
originating source then divergence of the already weakened light beam
would result in an extremely degraded power for interferometry, which
would actually make it impossible: only a few hundred photons per hour
would be received, a figure well below the shot noise in the light
sensing electronics (photodiodes).

To avoid this, \emph{active mirrors}, or \emph{transponders} are envisaged.
These are devices which, by means of a local oscillator, can capture phase
information of the incoming light beam and order the re-emission of the
full 1~watt laser with that phase information encoded in it. In this
form, interferometry with long arms is made possible.

\subsection{Orbit}

Obviously, there is no way one can possibly lock \lisa's interferometer
arm-length to 5 million km. In fact, the three spacecraft will be primarily
guided by the interplanetary gravitational field, i.e., the field of the
Sun and the other solar system bodies.

\begin{figure}[t!]
\centering
\includegraphics[width=12cm]{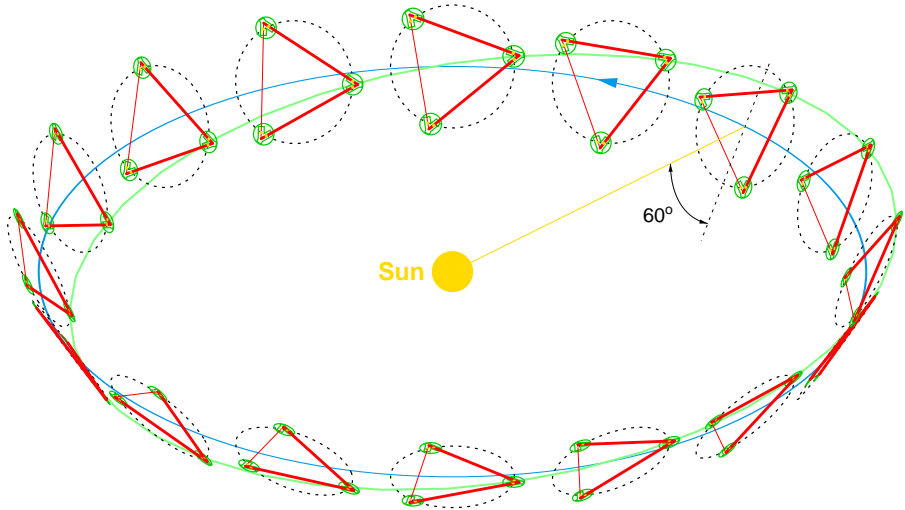}
\caption{\lisa's three spacecraft orbital evolution around the Sun. The
centre of the triangle follows the Earth's ecliptic, while the triangle
itself rotates clockwise around its centre once~per~year.
\label{Lobo_fig.3}}
\end{figure}

On the other hand, as we shall shortly see, \lisa will be required to
perform pico-metre interferometry, which means it should be able to detect
distance \emph{variations} between pairs of spacecraft to picometre precision.
This, in the milli-Hz frequency band.

The question naturally arises whether such measurement precision is
compatible with the laws of Celestial Mechanics which govern the motions
of the \lisa spacecraft. Astrometric studies~\menta{billf,Lobo_pete} have
shown that there exists an optimised satellite configuration, which is
displayed in figure~\ref{Lobo_fig.3}: the three spacecraft constellation
rotates clockwise around its barycentre with a period of one year, while
the barycentre follows the ecliptic (blue line in the drawing), 20$^\circ$
behind the Earth ---or, equivalently, 45 million km\footnote[1]{
This number is the result of an accepted compromise: if \lisa is
close to the Earth then gravitational perturbations distort its
configuration; if \lisa is far then we may run into telemetry problems.}.
The plane of the three satellites is inclined 60$^\circ$ relative to the
ecliptic, and the normal to it correspondingly describes a cone in the
sky. \lisa is thus not pointing to a fixed location. Each of the three
spacecraft is in an orbit which is very circular (eccentricity\,=\,0.01),
and the plane of each orbit is inclined 1$^\circ$ relative to the ecliptic.

\begin{figure}[b]
\centering
\includegraphics[width=14cm]{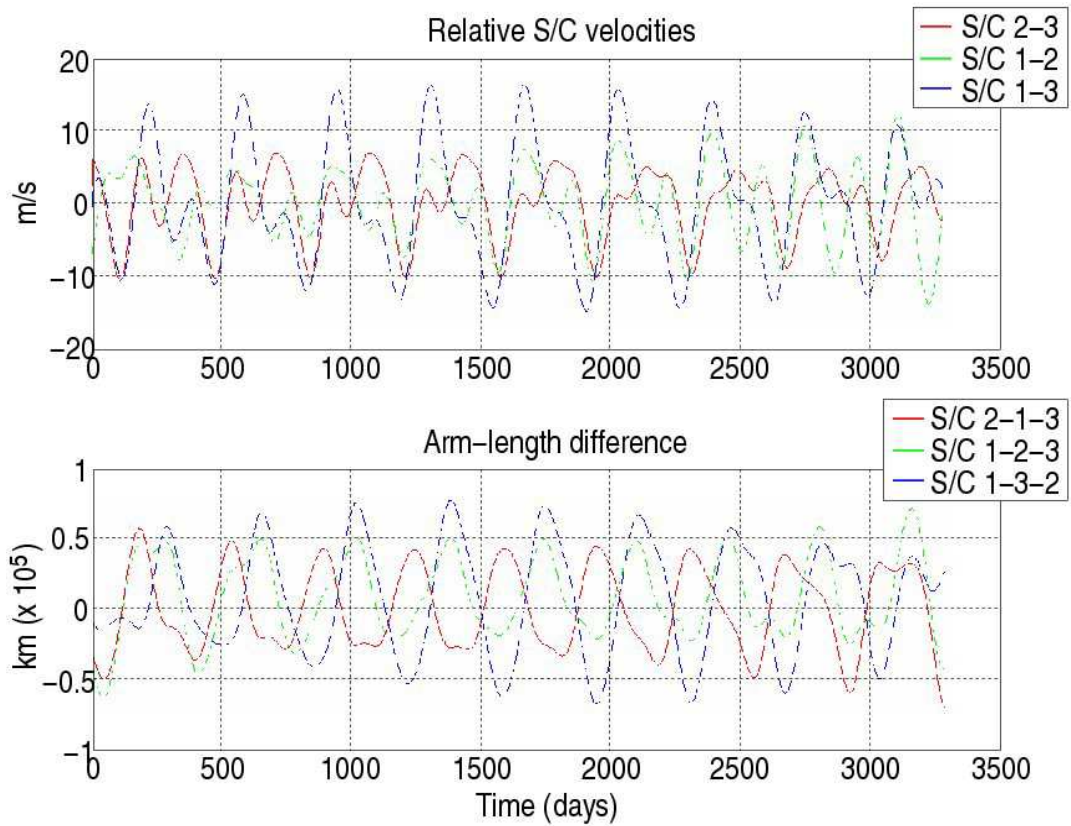}
\caption{Variations in relative velocity between spacecraft and differences
between arm-lengths over a period of 10 years, which is \lisa's extended
lifetime.
\label{Lobo_fig.4}}
\end{figure}

Even though this is basically the most stable orbit, in the sense that
it best keeps the triangular shape of the spacecraft constellation, such
shape is far from constant: the arm-lengths change, and so do the angles
between the arms; consequently, the spacecraft drift back and forth from
one another, too. These changes are not negligible, as we see in the
examples given in figure~\ref{Lobo_fig.4}: arm-length differences undergo
peak-to-peak variations of about 120\,000~km, and relative velocities can
be $\pm$15\,km/sec. Not shown in the figure, angles between contiguous
arms can vary by about 1~degree, peak-to-peak.

The reader may wonder how can one possibly do picometre interferometry in
a systems which drifts away by tens of thousands of kilometres. The answer
to this is that one is interested in length variations over time scales of
hours ---which correspond to milli-Hz frequencies. A look at the graphs
of figure~\ref{Lobo_fig.4} indicates that orbital changes happen instead
in time scales of months, i.e., far from the frequency of the GW signals
we are interested in. In addition, short time prediction of astronomical
ephemeris ---such as happen in the Solar System--- is very reliable, and
poses therefore no (theoretical) problem for detection.

\subsection{Time delay interferometry}

The varying length of the interferometer arms has however a setback of
a different nature. This is linked to the influence of \emph{frequency
fluctuations} on the readout of the phase meter, where light coming
from different arms is recombined to generate an interference pattern.

Traditional, equal arm-length Michelson interferometry does not have this
problem: light is generated in a laser source, then divided into two
beams at a beam splitter, then sent out to a pair of equally distant
mirrors, then reflected back by the latter and recombined again at
the beam splitter, then finally the interfering beam analysed by a
photo-detector, e.g.\ a photo-diode. A real laser does not have an
exact frequency, it actually \emph{fluctuates}. But the fluctuations
go undetected by the Michelson scheme: indeed, the interferometer
signal in this scheme is the phase \emph{difference} between the
two traveling beams, hence any fluctuations mutually cancel if
both beams have traveled the same distance from the instant of
split-up to that of recombination.

Phase noise thus enters the scenario whenever the optical path of the
laser beams differ. The intensity of noise depends on the stability of
the laser frequency: if the frequency fluctuates by $\delta f\/$ and the
arm-length difference is $\Delta L$, then the optical path difference
$\delta x\/$ is readily seen to be given (in first order) by
\begin{equation}
 \delta x = \frac{\delta f}{f}\,\Delta L
 \label{Lobo_eq.8}
\end{equation}

A very stable infra-red laser ($\lambda$\,=\,1.064\,$\mu$m) is
envisaged for \lisa, with a phase stability of one part in 10$^{13}$
per square root of Hertz, or $S^{1/2}_f$\,$\simeq$\,30\,Hz/$\sqrt{\rm Hz}$.
If we wish to do pico-metre interferometry ---see section~\ref{Lobo_sec.3}
below--- the length variations we can tolerate are, according to
formula~(\ref{Lobo_eq.8}), $\Delta L\/$\,$<$\,200\,metres. But, as we have
just seen, \lisa undergoes much larger changes, so we have a problem here.

This is precisely the problem Time Delay Interferometry, or TDI, addresses.
TDI is a \emph{post-processing} technique to remove frequency noise from
interferometer data. It is therefore not implemented in hardware, but
purely in software. An excellent and updated review will be found
in~\menta{tdi}. Here we shall only give the simplest example in order
to illustrate the concept.

\begin{figure}[t!]
\centering
\includegraphics[width=13cm]{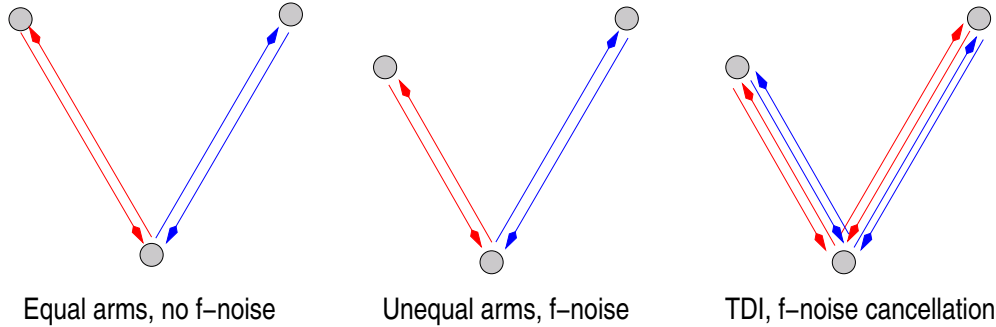}
\caption{Concept of the TDI variable of equation~(\protect\ref{Lobo_eq.10}):
frequency noise is proportional to the difference between the blue and the
red beam lengths.
\label{Lobo_fig.45}}
\end{figure}

Assume $\phi(t)$ is the phase of the laser as it enters the beam splitter
before the light is distributed to two of the \lisa arms, which have
lengths $L_1$ and $L_2$, respectively. Let $y_1(t)$ and $y_2(t)$ be the
phase readings at the beam splitter \emph{after} the light has come back
from each of the arms. Then
\begin{eqnarray}
 y_1(t) & = & \phi(t-2L_1/c) - \phi(t) + \chi_1(t) + n_1(t)
 \label{Lobo_eq.9a} \\
 y_2(t) & = & \phi(t-2L_2/c) - \phi(t) + \chi_2(t) + n_2(t)
 \label{Lobo_eq.9b}
\end{eqnarray}
where $\chi_1(t)$, $\chi_2(t)$ are GW phase shifts, and $n_1(t)$, $n_2(t)$
are other noise contributions, not due to frequency random jitter. This
latter term is included in $\phi(t)$. The following TDI variable can now
be defined:
\begin{equation}
 X(t) = [y_1(t) - y_2(t)] - [y_1(t-2L_2/c) - y_2(t-2L_1/c)]
 \label{Lobo_eq.10}
\end{equation}
where time delays have been adequately chosen to cancel out frequency
noise effects. This is of course based on the fact that laser
frequency fluctuations enter into $y_1(t)-y_2(t)$ and into
$y_1(t-2L_2/c)-y_2(t-2L_1/c)$ with the same time dependence structure.
Figure~\ref{Lobo_fig.45} shows a conceptual scheme of how this TDI
variable cancels frequency noise.

There are many other TDI variables which cancel various noise components.
The common philosophy is that they are generated by linear combinations
of suitably time-shifted interferometer readouts.

\section{\textsl{LISA}'s scientific objectives and requirements}
\label{Lobo_sec.3}

\begin{figure}[b!]
\centering
\includegraphics[width=11cm]{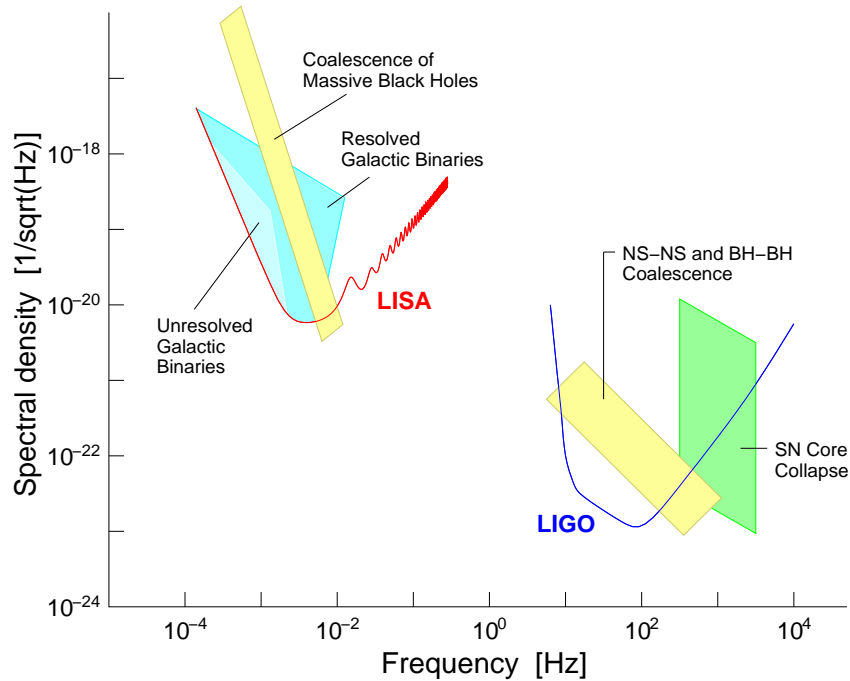}
\caption{Spectral density of noise for \lisa. Ground based \textsl{LIGO}
sensitivity is also plotted for comparison. A few reference GW sources
are quoted, too.
\label{Lobo_fig.5}}
\end{figure}

The scientific objectives of \lisa and the mission requirements are
tightly bound together: there are noise constraints which limit what
can be achieved in practice, and there are GW sources which one aims
to see as best as possible. At the time of writing there is still
debate within the \textsl{LIST} (\lisa International Science
Team~\menta{list}) on fine structure figures. Approximately, the
sensitivity curve for \lisa is shown in figure~\ref{Lobo_fig.5}.

According to this curve, the strain sensitivity is
$\sim$7$\times$10$^{-21}$\,Hz$^{-1/2}$ at 3 mHz, where the noise
minimum is located. Sensitivity worsens in a V-shaped curve as we
move out of the trough in frequency domain. In terms of displacement
noise, and taking into account that GW strain is a measure of
\emph{relative} distance changes, $h\/$\,=\,$2\delta L/L$~\menta{wein},
the above is equivalent to
$S^{1/2}_L$\,$\simeq$\,20~pico-metres/$\sqrt{\rm Hz}$ at 3 mHz.

As already stressed in the Introduction section, many and important
GW sources are expected in \lisa's low frequency band. GW Sighting
and characterisation of such sources are therefore part of the mission's
scientific objectives. While objectives of this kind strongly rely on
our belief that General Relativity is correct ---in so far as we make
\emph{quantitative} estimates of GW intensities---, we may not forget
that observations of the Universe through the \emph{gravitational
window} have never been made so far. So, when \lisa gets operative,
the unexpected should be expected\ldots\ The GW Universe could well
provide new fresh evidence that some of our current astrophysical
and/or cosmological views need changes, or even hint at, or directly
solve such difficult and paradoxical problems as modern Cosmology
faces. By way of speculation, one may for example recall that 22\,\%
of the Universe is made of dark matter~\menta{wmap}; dark matter is
not electromagnetically visible yet it does indeed gravitate ---hence
GWs could be generated there and bring information of this largely
unknown form of matter.

There are three major areas of Physics where we can identify potential
contributions by \lisa. We list them below, along with some conjectured
sources:

\begin{enumerate}
\item {\sf Astronomy and Astrophysics:}
 \begin{enumerate}
 \item Are Kerr black holes at the centre of active galactic nuclei?
 \item How do super-massive black holes form and grow?
 \item We can observe the evolution of thousands of galactic binaries
 \end{enumerate}
\item {\sf Cosmology:}
 \begin{enumerate}
 \item How do super-massive black holes interact and contribute to the
  formation of galaxies?
 \item Study massive black holes up to redshifts of $z\/$\,=\,30
 \item Observe high redshift objects at gravitationally calibrated
  distances to more precisely determine the Hubble constant
 \item Independent measurements of dark energy parameters
 \end{enumerate}
\item {\sf Fundamental physics and discovery:}
 \begin{enumerate}
 \item Test General Relativity at its \emph{strong field} limit by mapping
  space-time near black holes
 \item Compare white dwarf binaries to theory (calibration binaries)
 \item Evidence of gravitational dynamics on cosmological scales:
 \begin{itemize}
  \item Is there a first order phase transition at TeV scale energies?
  \item Are there extra dimensions at the sub-millimeter scale?
  \item Do cosmic superstrings exist?
 \end{itemize}
 \item There will likely be so far unforeseen phenomena, unveiled by
  GW observations
 \end{enumerate}
\end{enumerate}

To build hardware and software capable of implementing the requirements
set forth in figure~\ref{Lobo_fig.5} is no mean feat, from the technology
point of view. Actually, the technological road to \lisa is by no means
an easy one: the system is complex, and has many subsystems which will
ultimately be integrated into a coherent whole. In fact in \lisa the
traditional division between spacecraft and payload cannot be clearly
drawn: the payload takes continuous action on spacecraft navigation
decisions, while the spacecraft computer continuously overviews and
authorises actions by the payload computer.

\section{The \textsl{LISA} payload}

We shall not delve into the detailed structure and assembly of the various
parts of \lisa. We shall however give a brief overview of its two most
important subsystems ---from the conceptual viewpoint---, i.e., the
\emph{drag free} and the Metrology subsystems.

\subsection{The \textsl{drag free} subsystem}
\label{Lobo_sec.41}

Detection of GWs with \lisa's is only possible if the test masses follow
nominally geodesic trajectories, i.e., those defined by the interplanetary
gravitational field, \emph{plus} the GWs themselves. These show up as
time dependent gravity gradients which can be identified above an otherwise
stationary background by certain specific signatures.

A most delicate problem in \lisa therefore is to make sure that the test
masses do actually move along geodesics to \emph{very high precision}.
The problem to ensure this is that interplanetary space is in fact a
considerably hostile medium: solar radiation pressure, ionising particle
fluxes, and environmental magnetic fields are among the agents which
would perturb geodesic motion should the test masses be floating
\emph{unshielded} in their orbits.

Apart from hosting the measuring and control instrumentation, the
spacecraft play a fundamental role in providing the test masses adequate
protection against external agents. Test masses are freely floating
\emph{inside} the spacecraft, and it is therefore the latter which
receive the impact of perturbations, eventually being driven away from
their geodesics. In order not to drag along the test masses with them, a
so called \emph{drag free} system is implemented in the satellites: this
consists in a \emph{gravitational reference sensor}, and an associated
\emph{actuation system}.

\begin{figure}[t]
\centering
\includegraphics[width=13cm]{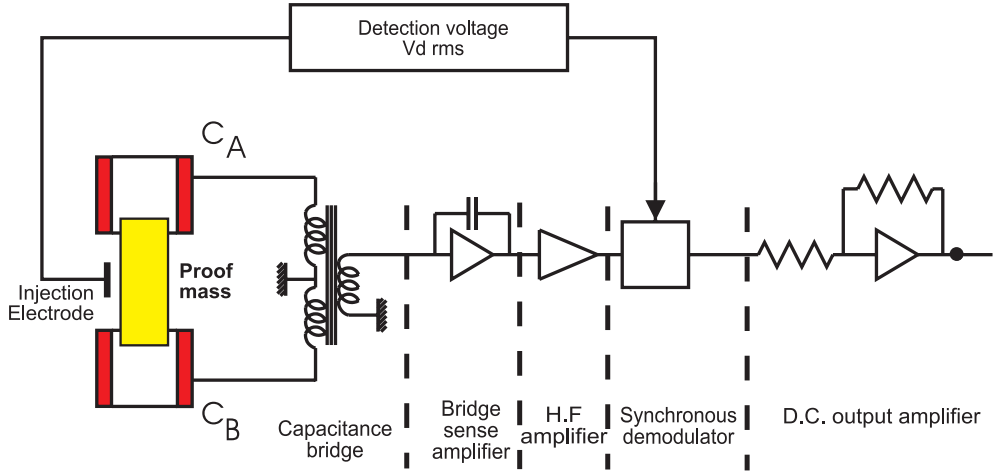}
\caption{Scheme of the GRS capacitive sensing, only one channel.
\label{Lobo_fig.6}}
\end{figure}

Each test mass is housed in a box whose walls are metallic plates which
form capacitors with the faces of the test mass itself. In equilibrium
conditions the mass is centred in the housing, and deviations thereof
result in capacity changes, which are detected by corresponding bias
voltage variations~\menta{billdf}, see figure~\ref{Lobo_fig.6}. This is
called \emph{gravitational reference sensor}~(GRS). Its error signal is
a voltage, proportional to the difference of the gaps on opposite sides
of the test mass, and is used to send suitable ignition commands to a
set of micro-thrusters which restore the centred positions of the test
masses by acting on the spacecraft only. Micro-thrusters produce very
gentle micro-newton forces by the ejection of ions, or atoms.

The electrode housing actually has several electrodes per face of
the cubic test mass, whereby it can obtain and process full attitude
control of the spacecraft ---the so called DFACS, or Drag Free and
Attitude Control System. The GRS, together with its complex electronics
and control software, is a most essential part of \lisa. In a system
with so many degrees of freedom, and where \emph{two} test masses are
hosted in the same spacecraft, full implementation of the DFACS is
not only subtle, it also requires eventual \emph{back actuation} on
the test masses by the GRS itself, though perpendicular to the
corresponding interferometer arm.

\subsection{Optical Metrology System in \textsl{LISA}}

The interferometry in \lisa has some differences with that in earth based
GW detectors. The latter are essentially Michelson. Apart from the already
commented fact that \emph{active mirrors} are needed in \lisa to compensate
for energy loss down the very long arm-length (with a small laser power of
1~watt), the system works in a three-stage scheme: first, the position of
the test mass relative to the spacecraft is established; second, the
distance between spacecraft is determined; and third, the position of
the other end mass relative to its spacecraft is measured. A heterodyne
operation mode is accepted in all cases~\menta{lfr}.

The optical metrology consists of:
\begin{itemize}
 \item The laser assembly, which includes the laser source proper, a
  Yag-Neodymium, 1~watt, 1.064~micron laser, plus the acousto-optical
  modulators (for heterodyning) and various stability and feedback
  controls. There is one per spacecraft (plus redundancy).
 \item The optical bench, which incorporates mirrors, beam splitters
  and photo-diodes for all the required interferometry modes. There is
  one per test mass, so two per spacecraft. No redundancy here.
 \item The telescope, which is a Cassegrain with a primary and a
  secondary mirror. The telescope collects the light coming from
  the remote spacecraft, and also re-emits it, sending the light
  in exactly opposite direction. There is one telescope per test
  mass, hence two per spacecraft. No redundancy, either.
\end{itemize}

\begin{figure}[t]
\centering
\includegraphics[width=12cm]{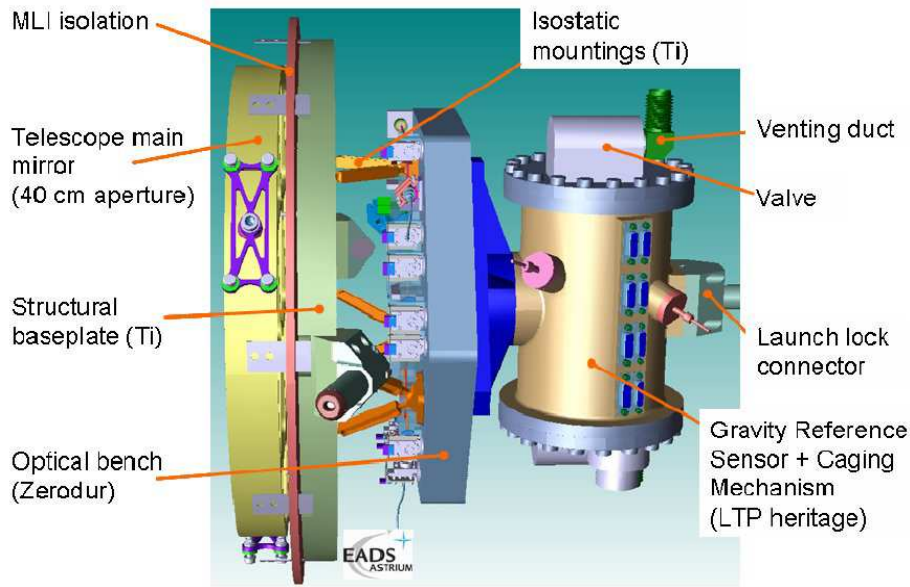}
\caption{Mounting of the GRS on the Optical Bench. The telescope's
primary mirror is visible on the left. Thanks are due to to Peter
Gath and Ulrich Johann, from EADS-Astrium Friedrichshafen for
providing this picture.
\label{Lobo_fig.7}}
\end{figure}

Figure~\ref{Lobo_fig.7} represents a design view of the GRS mounting,
together with its optical bench and telescope for a single test mass.

\section{\textsl{LISA PathFinder}}

The reader who has made it this far has surely acquired a flavour of the
practical difficulties and challenges the \lisa designers, engineers and
even scientists have to face. Ultimately, it is all of course due to the
extreme weakness of the GW signal we want to detect.

While the Science case to fly \lisa is widely considered a strong one
within the community, it is an obvious matter of course that the
mission should  only be launched when solid expectations that it
will work are in place. In the space expert jargon, the question of
\emph{technology readiness} (or maturity) is the one which must be
fixed before the mission gets full approval to take off.

Many of the technological capacities needed for \lisa can be developed
and tested on ground: an instrument which works in the laboratory can,
in most cases, be tweaked in such a way that it will work in space, too.
This requires the use of radiation-hard materials and electronic components,
plus a number of structural and vibration tests which constitute the
process known as \emph{space qualification} procedure. Such process is
very well established by space agencies, based on experience of about a
half century of space flight, and resources to qualify a laboratory
prototype according to the appropriate protocols are abundant in the
specialised market.

But \lisa's \emph{drag free} subsystem, which is the core of the mission,
cannot be tested on earth: the extremely precise and long duration free
fall conditions needed for a test of the GRS are simply orders of magnitude
beyond the best gravity--free laboratory available on ground. In view of
this, and in order to mitigate the risks of a direct launch of \lisa, which
is very expensive indeed, the European Space Agency (\esa) has decided to
first fly a precursor mission where \lisa technologies are put to test.
The results of this mission will of course heavily bear on \lisa's timing
and final design.

The precursor mission is called \lisa \textsl{PathFinder} (\lpf). It is
part of \esa's \textsl{SMART} (Small Missions for Advanced Research
Technology) programme, of which it will be the \emph{second} mission.
\lpf is scheduled to fly in 2009, and is now entering its industrial
implementation phase, therefore a very advanced state of development.
Seven European countries (voluntarily) participate in \lisa
\textsl{PathFinder}, which means national funding is provided in each
case for the share of responsibility accepted by the corresponding party
within the mission consortium. Specific national funding covers about
20\,\% of the total mission cost, the rest being provided by \esa.
The seven countries are: Italy, Germany, United Kingdom, Spain, Holland,
Switzerland, and~France.

The payload aboard \lpf is the so called \lisa Technology Package (\ltp),
whose details we shall come to shortly. The \ltp is an entirely European
instrument. However, given that \lisa is conceived as a joint venture
between \esa and \nasa, it was long scheduled that \nasa would provide
their own version of the \lisa precursor technology, to fly with the
same space platform as the \ltp. The American contribution was called
Disturbance Reduction System (DRS). Unexpected difficulties during the
development and test of the DRS have unfortunately led \nasa to very
significantly descope their contribution to the PathFinder test, which
will now essentially consist of the \ltp. \nasa will however still provide
a set of alternative micro-thrusters, together with software to drive them
on the basis of data handed by the European Gravitational Reference Sensors.

In the following sections we describe the main parts and functionalities
of \lpf and the \ltp, as well as their motivation and purposes.

\section{\textsl{LPF} mission description}

\lpf will consists in a single spacecraft, hosting a pair of test masses.
One of them acts as the GRS, while the other will be acted according to
precisely defined operation modes. A laser interferometer will check that
the level of \emph{drag free} is the one required.

\subsection{Orbit and general mission details}

\begin{figure}[t]
\centering
\includegraphics[width=7.5cm]{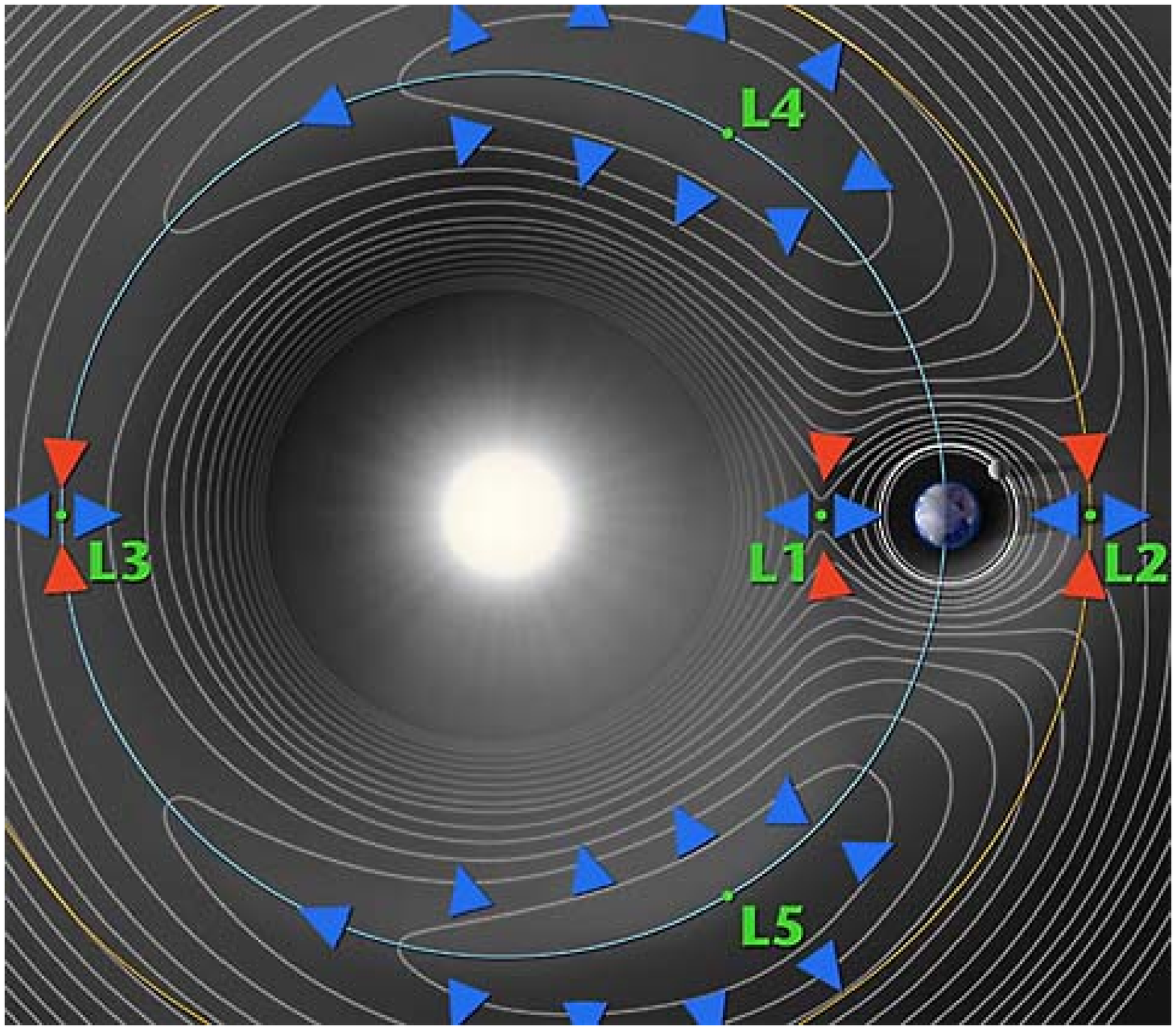}\quad
\includegraphics[width=7cm]{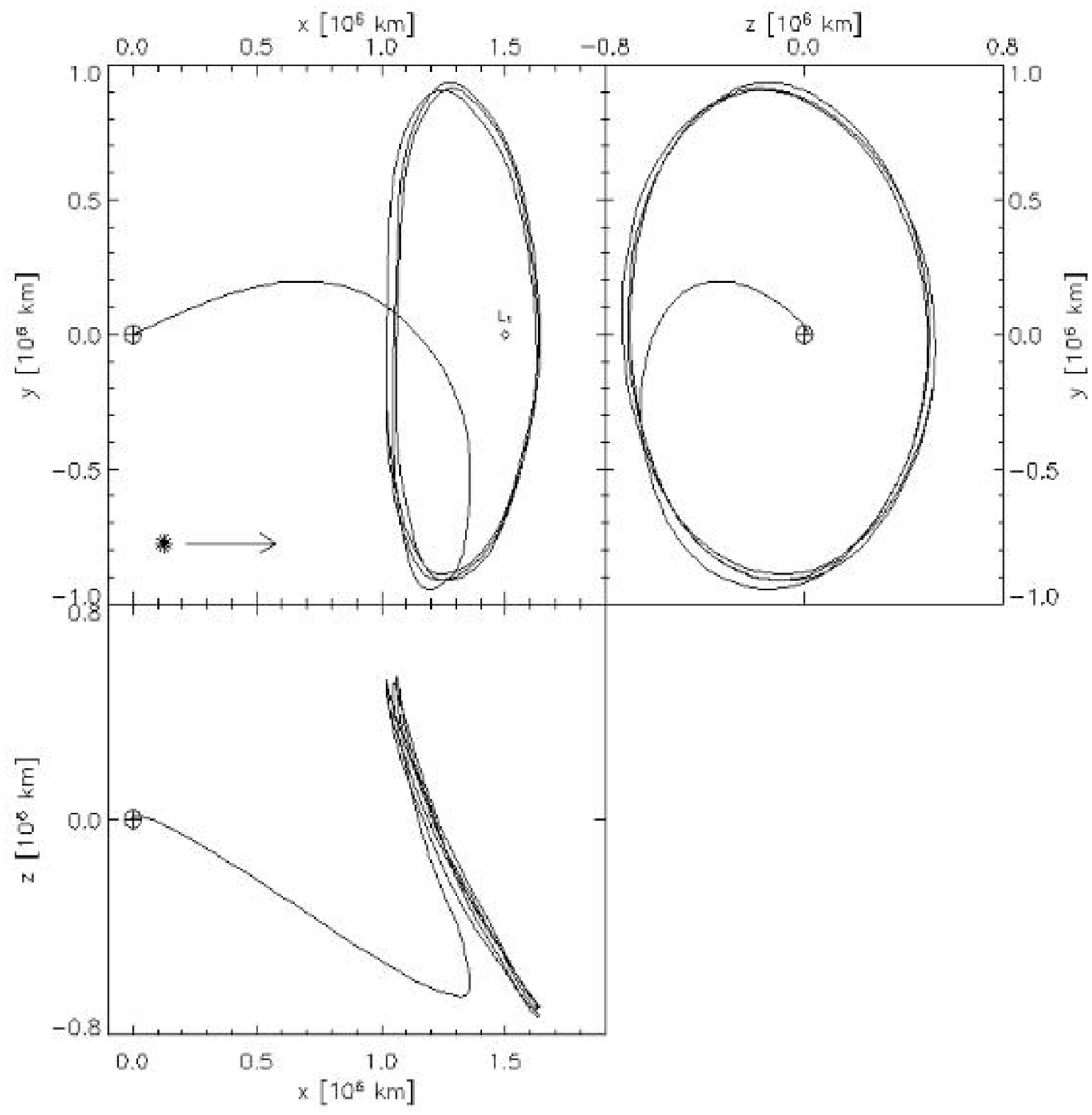}
\caption{Left: the Lagrange points of the Earth-Sun system, and the
gravitational field lines (the plot is \emph{not} to scale). Right:
Lissajous orbit for \lpf, from the the three coordinate planes. Axes
are labeled in millions of kilometres.
\label{Lobo_fig.8}}
\end{figure}

\lpf will operate from the Earth-Sun Lagrange point $L1$ ---see
figure~\ref{Lobo_fig.8}, left. As we see indicated in the figure, $L1$
is a \emph{metastable} position, hence the spacecraft will actually
revolve around it in a Lissajous orbit: this is a nearly circularly
shaped trajectory in a plane which is almost perpendicular to the
Sun-Earth line, see figure~\ref{Lobo_fig.8}, right. Its diameter is
almost 2 million kilometers, and it takes about 6 months to perform
a complete rotation around it.

\lpf will be launched in a Rockot vehicle, from the Russian space field
at Pliesietsk, near Arkangelisk in Northeastern Russia. Launch date is
currently fixed for the third quarter of 2009. The vehicle will be
initially inserted into a slightly elliptic low Earth orbit, with an
inclination of~63$^\circ$. The orientation of the line of apsides of
the parking orbit must be adjusted to target for an operational orbit
that fulfills the station visibility constraints, which is a minimum of
8~hours of visibility from the ground-station in Villafranca del Castillo,
near Madrid (Spain). A sequence of between 11 and 15 manoeuvres will bring
the spacecraft to a transfer towards $L1$, which is 1.5 million kilometres
from Earth. In each manoeuvre, a speed increment of some 3~km/sec is
applied to the spacecraft at the orbit's perigee. The new orbit after
each thrust maintains the perigee altitude, while the apogee gets further
away, thus progressively increasing the orbital eccentricity until the
spacecraft gets eventually detached.

The trip time to operational orbit is some three months. The minimum
mission lifetime is 200 days, with possible extensions if deemed necessary
by the mission international consortium, and by \esa.

\subsection{\textsl{LTP} concept and top level requirements}

\begin{figure}[b!]
\centering
\includegraphics[width=7cm]{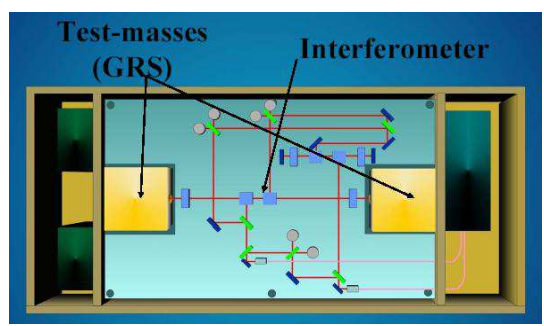}
\caption{Conceptual diagramme of the \ltp.
\label{Lobo_fig.9}}
\end{figure}

The \lpf mission is intended to test in flight a number of essential parts
of \lisa, most notably the \emph{drag free} subsystem, but also picometre
interferometry in space and other subsystems and software. For that, a
\lisa arm is squeezed to 30 cm, as shown in figure~\ref{Lobo_fig.9}.

It has become common practice to convert GW noise spectral densities into
\emph{acceleration} noise spectral densities, which seems to be a more
manageable concept from the experimentalist's point of view. To make the
translation is however quite simple.

Let us first of all recall from the first paragraph of
section~\ref{Lobo_sec.41} that GW effects show up as gravity gradients,
or \emph{tides}. Equivalently, by relative accelerations ---hence
forces--- in the language of Classical Mechanics, or \emph{geodesic
deviations}, as General Relativity experts normally prefer. Taking
as reference one of the test masses, the \emph{relative} acceleration
of the other is given by
\begin{equation}
 \Delta a\equiv\frac{d^2\Delta x}{d t^2} = \frac{L}{2}\,\frac{d^2h}{dt^2}
 + \frac{\Delta F}{m}
 \label{Lobo_eq.11}
\end{equation}
where $m\/$ is the mass of the second TM, $\Delta x\/$ is the relative
displacement, $t\/$ is the quasi-Lorentzian time coordinate, and $F\/$
embraces all non-gravitational forces, such as thermal, magnetic, electric,
etc., and also non-inertial forces such as e.g.\ rotations. It is expedient
to rewrite this expression in frequency domain:
\begin{equation}
 \widetilde{\Delta a}(\omega) = -\omega^2\,\widetilde{\Delta x}(\omega) =
 -\frac{L}{2}\,\omega^2\,\widetilde{h}(\omega)
 + \frac{\widetilde{\Delta F}(\omega)}{m}
 \label{Lobo_eq.12}
\end{equation}
where a \emph{tilde} (\,$\tilde{}$\,) stands for Fourier transform. If
we now take spectral densities in the last equation we find, assuming
of course that the true GW signal is deterministic, that the spurious
forces $\Delta F$ actually \emph{fake} a GW noise with equivalent rms
spectral density
\begin{equation}
 S_h^{1/2}(\omega) = \frac{2S_{\Delta F}^{1/2}(\omega)}{mL\omega^2}\ ,
 \quad {\rm equivalent\ noise}
 \label{Lobo_eq.14}
\end{equation}

In terms of acceleration noise, the sensitivity requirement for \lisa
can be recast in the form
\begin{equation}
 S^{1/2}_a(\omega)\leq 3\times10^{-15}\,\left[
 1 + \left(\frac{\omega/2\pi}{3\ {\rm mHz}}\right)^{\!\!2}\right]\ 
 \frac{\rm m}{{\rm s}^2\,\sqrt{\rm Hz}}
 \ ,\quad 0.1\ {\rm mHz}\leq\frac{\omega}{2\pi}\leq
 0.1\ {\rm Hz}\,,\quad\lisa
 \label{Lobo_eq.15}
\end{equation}
where the quadratic dependence on frequency here is simply because,
according to equation~(\ref{Lobo_eq.14}), the acceleration noise can
increase like $\omega^2$ at high frequencies without affecting the GW
sensitivity.

It has been agreed that performance of \ltp can be considered fully
satisfactory if the requirement for \lisa is relaxed by one order
of magnitude, both in spectral density of noise and in frequency
band. More specifically~\menta{tlsr},
\begin{equation}
 S^{1/2}_a(\omega)\leq 3\times10^{-14}\,\left[
 1 + \left(\frac{\omega/2\pi}{3\ {\rm mHz}}\right)^{\!\!2}\right]\ 
 \frac{\rm m}{{\rm s}^2\,\sqrt{\rm Hz}}
 \ ,\quad 1\ {\rm mHz}\leq\frac{\omega}{2\pi}\leq
 30\ {\rm mHz}\,,\quad\ltp
 \label{Lobo_eq.16}
\end{equation}

The reason for this order of magnitude margin is this: if we could
directly aim at \lisa's requirements then it would not make too much
sense to fly a precursor mission, which costs time and money. On the
other hand, the best \emph{drag free} ever flown to date is several
orders of magnitude less demanding than what we are requiring just
for the \textsl{PathFinder}, so too big a jump might be unrealistic.
If \lpf returns satisfactory results, then we will have good hints
on how to improve it so as to make it to \lisa. We shall come back
to this important issue.

\subsection{Philosophy of the \textsl{LTP} test}

The instrumentation and design of the \ltp must ensure that any
\emph{residual accelerations}, i.e., those of unknown physical origin,
be below the requirement expressed by equation~(\ref{Lobo_eq.16}). This
requires in turn a detailed \emph{apportioning} of different
contributions to the background noise.

An essential fact in this respect is the following: certain perturbing
agents \emph{couple} the spacecraft structure to the test masses, while
others are independent of them. A sort of master equation can thus be
set up:
\begin{equation}
 a_{\rm noise} = \frac{F_{\rm int}}{m} +
 \omega_{\rm p}^2\,\underbrace{\left(x_{\rm n} +
 \frac{F_{\rm S/C}}{M\omega_{\rm fb}^2}
 \right)}_{\mbox{\scriptsize\parbox[t]{22.5 ex}{
 $x_{\rm S/C}={\rm S/C}\rightleftharpoons{\rm TM}$
 {\rm relative\ distance}}}}
 \label{Lobo_eq.17}
\end{equation}
where $F_{\rm int}$ is the random force acting on a given test mass,
$\omega_{\rm p}^2$ is the elastic constant (or \emph{stiffness}) of
coupling between the test mass and the spacecraft (can be a negative
number), $x_{\rm n}$ is the random displacement fluctuations of the
test mass and $F_{\rm S/C}$ the random force fluctuations acting on
the spacecraft, $\omega_{\rm fb}^2$ is the elastic constant of the
coupling between $F_{\rm S/C}$ and the test mass, related to the
response time of the actuation system. Finally, $M\/$ is the
spacecraft mass, and $m\/$ is the test mass mass.

All the parameters in~(\ref{Lobo_eq.17}) must be evaluated on the basis
of experimental measurement, and to this end several measuring runs and
operation modes have been designed~\menta{master}. In each case, the
\ltp interferometer will be used as a \emph{diagnostic} instrument, as
its foreseen sensitivity is already sufficient~\menta{gh} to meet the
measurement demands necessary to establish the validity of the limit
set by equation~(\ref{Lobo_eq.16}).

\subsection{\textsl{LTP} functional architecture}

\begin{figure}[t]
\centering
\includegraphics[width=15cm]{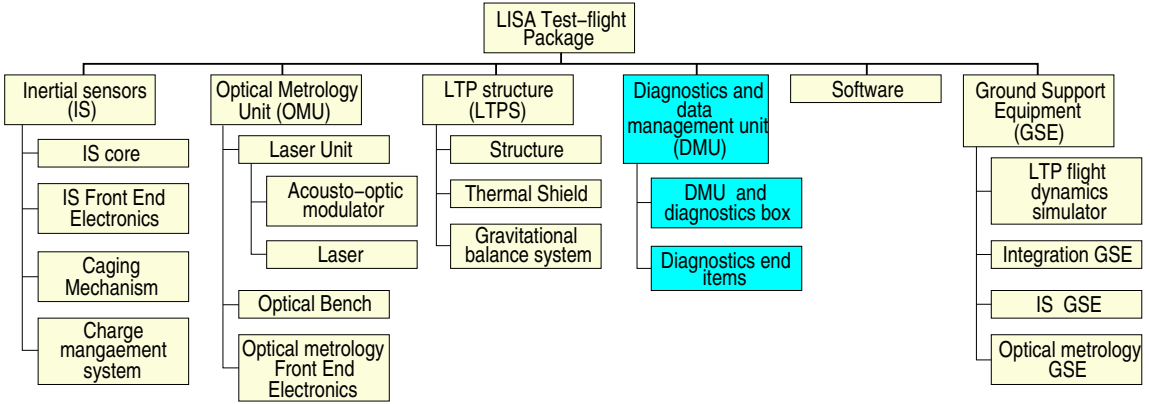}
\caption{\ltp functional diagramme scheme. Highlighted, the Spanish
contribution.
\label{Lobo_fig.10}}
\end{figure}

It is most clearly explained graphically, as in figure~\ref{Lobo_fig.10},
where all the \ltp parts are ordered by function. We shall not attempt
to give even a summary description of each subsystem here. We shall
instead concentrate on the diagnostics and \dmu subsystem, which is
the highlighted branch of the drawing. Such will be the subject of
the next sections.

\section{The Data and Diagnostics Subsystem, or DDS}

As shown in the highlighted branch of figure~\ref{Lobo_fig.10}, it
consists of two main parts: the Data Management Unit, or \dmu, and the
Diagnostics Elements. We first enumerate their components:

\begin{enumerate}
 \item {\sf DMU.} In spite of its name, it is the \ltp computer. It is
  responsible for driving and control of the diagnostics items, and for
  such important tasks as the acquisition and on-board processing of
  phasemeter data. Some of these data are the mission science data.
  The \dmu also interfaces with other systems, in particular with the
  main mission computer ---so called OBC (On Board Computer). The \dmu
  has three main components, all of them \emph{duplicated} as a resource
  against failure or malfunction. They are:
 \begin{enumerate}
  \item Power Distribution Unit (PDU)
  \item Data Acquisition Unit (DAU)
  \item Data Processing Unit (DPU)
 \end{enumerate}
 \item {\sf Diagnostics Elements.} These are a number of sensors, plus
  their associated electronics, which are intended to monitor various
  disturbances happening inside the \ltp. They are the following:
 \begin{enumerate}
  \item Precision temperature sensors. There are {\bf 22}, distributed
   across the \ltp
  \item Calibration heaters. There are {\bf 14}, distributed in the
   GRS, the Optical Window and the suspension struts
  \item Precision magnetometers. There are {\bf 4} of \emph{fluxgate}
   type, surrounding the \ltp
  \item Calibration coils. There are {\bf 2}, one per GRS and both
   aligned with the line joining the test masses
  \item Radiation Monitor. There is just one, attached to the spacecraft's
   shear walls.
 \end{enumerate}
\end{enumerate}

Before going into any detail on the above, let us first discuss the
relevance of the DDS for the \textsl{PathFinder} mission. While it
is fairly obvious that the \dmu is mission critical ---i.e., if there
is no \dmu then there is no mission---, the reasons for the need of
the Diagnostics Elements are less trivial yet fundamental for the real
motivation of the \lpf: to pave the way to jump across the sensitivity
gap between \lpf and \lisa.

\subsection{Why are the Diagnostics Elements crucial for \textsl{LPF}?}

To answer this question, we first need to ask another question. Imagine
\lpf works perfectly, i.e., it proves to be compliant with the top level
requirements expressed by equation~(\ref{Lobo_eq.16}) to exquisite precision.
While this will likely bring to the \lpf community a truly rewarding
sense of achievement, a very important question will still remain open
after such potential success. It is: ``OK, but this an order of magnitude
less sensitive than \lisa; \emph{how do we reach there?}''.

There are two tasks here: one is to identify which parts of the \ltp are
liable to sufficient improvement, the other to pursue suitable research
to actually improve the system. The Diagnostics Elements constitute the
essential tool for the first task, and can also be of great help for
the second. Let us see now how this comes about.

Recall first of all that the \ltp science readout is provided
by the \emph{phasemeter}, i.e., the interferometer output. There are
several interferometry output channels ---some of them for diagnostics,
too---, but there is only one \emph{science channel} proper. Relative
accelerations can be easily translated into displacements (in Fourier
domain there is just a $-\omega^2$ factor between them), thence into
phase shifts. The \ltp phase readout is expected to be pure noise, no
signals whatsoever (up to some serendipitous unpredictable discovery),
with a spectral density at or below top level requirement values.

The problem we want to solve is to identify the magnitude of each of
the various sources of noise which ultimately combine into the measured
phase noise, where they add together entangled and undifferentiated.
For this, we have the diagnostics elements.

Notice, however, that sensors alone are not sufficient: a question
remains which is the relationship between, say, the temperature field
across the \ltp and the phase readout. To establish that relationship
by theoretical analysis of the spacecraft structure is basically hopeless
---there are too many parts: optical elements, flanges, screws,
connectors, brackets, glass, harness,\ldots\ It is more practical
to resort to \emph{in situ} measurements by a finite number of
sensors, and to devise a procedure to generate controlled intense
signals (for example thermal shocks), whose effect be clearly seen
in the phasemeter. If temperature measurements are made simultaneously
with the shocks, the above procedure provides a way to establish the
sought for relationship.

We can quantify the above. Let $\phi(t)$ be the phase readout, which
is a sampled time series. This total phase can be \emph{apportioned}
into various contributions ---magnetic, thermal, laser phase noise,
etc. Noting generically by $\alpha$ any one of these, we can write
\begin{equation}
 \tilde{\phi}(\omega) = \sum_\alpha\,\tilde{\phi}(\omega;\alpha)
 \label{Lobo_eq.18}
\end{equation}
where $\tilde{\phi}(\omega;\alpha)$ is the contribution of the $\alpha$
agent to the total $\tilde\phi(\omega)$, and where Fourier domain
magnitudes have been used, as they are the ones which will be used
in normal practice.

Our concern is the determination of the \emph{feedthrough} coefficients
\begin{equation}
 H(\omega;\alpha) = \frac{\partial\tilde\phi(\omega)}{\partial\alpha} =
 \frac{\partial\tilde\phi(\omega;\alpha)}{\partial\alpha}
 \label{Lobo_eq.19}
\end{equation}

In the linear regime, knowledge of the feedthrough coefficients suffices
to calculate the contribution $\tilde{\phi}(\omega;\alpha)$:
\begin{equation}
 \tilde{\phi}(\omega;\alpha)\simeq H(\omega;\alpha)\,\alpha
 \label{Lobo_eq.20}
\end{equation}

The problem therefore translates into how to actually measure
$H(\omega;\alpha)$. The idea is to generate in the system such
a strong $\alpha$-signal that the readout gets dominated by
$\tilde\phi(\omega;\alpha)$. In this circumstance we have
$\tilde\phi(\omega)$\,$\simeq$\,$\tilde\phi(\omega;\alpha)$,
and hence a series of measurements of phase for various values
of $\alpha$ results in an estimate of $H(\omega;\alpha)$.
Accuracy of the estimate is of course dependent on signal-to-noise
ratio of the applied control signal. For the \ltp, this is required
to be 50 or larger~\menta{ddssr}.

Summing up, we \emph{need} diagnostics sensors, plus suitable controlled
perturbation generators, if we want to be able to focus future research
activities towards \lisa. Such is of course our ultimate motivation to
build and fly \lpf. We devote the following sections to review the
research done so far on diagnostics elements for the \ltp. \dmu
develpment will be left out, due reasons of~space.

\section{Thermal diagnostics}

Temperature fluctuations are a source of disturbances in the \ltp.
Thermal gradient fluctuations cause pressure differences between
test mass faces which tend to push them away from their centred
position. This happens due to \emph{radiometer} effect, \emph{radiation
pressure} effect, and \emph{asymmetric outgassing}. In addition,
temperature fluctuations do also affect the refractive index of
optical elements in the optical bench, both due to temperature
dependence of this quantity, and to stress induced dependence.

All in all, temperature fluctuations must be maintained below a certain
level, otherwise the \ltp readout noise will grow unacceptably high.
Temperature fluctuation noise is not the only source of noise in the
\ltp, and a limit has been set to 10\,\% of the top level requirement,
equation~(\ref{Lobo_eq.16}). Analysis shows~\menta{nos1} that
temperature fluctuations should be maintained at
\begin{equation}
 S_{T}^{1/2}(\omega)\leq 10^{-4}\,{\rm K}/\sqrt{\rm Hz}\ ,
 \quad 1\,{\rm mHz}\leq \omega/2\pi \leq 30\,{\rm mHz}
 \label{Lobo_eq.21}
\end{equation}

This basic stability requirement sets the reference for thermal diagnostics,
both for temperature sensor and calibration heaters performance.

\subsection{Temperature sensing}

The Industrial Architect is responsible to ensure that the spacecraft
is able to maintain the \ltp temperature conditions compliant with
equation~(\ref{Lobo_eq.21}). If we are to measure temperature fluctuations
below that level, our sensing system must of course be less noisy. We
have defined that one order of magnitude margin should be given to
sensing, or
\begin{equation}
 S_{T,\ {\rm measurement}}^{1/2}(\omega)\leq 10^{-5}\,{\rm K}/\sqrt{\rm Hz}\ ,
 \quad 1\,{\rm mHz}\leq \omega/2\pi \leq 30\,{\rm mHz}
 \label{Lobo_eq.22}
\end{equation}
is the temperature sensing system requirement.

Suitable sensors can be either platinum resistors (PTD) or thermistors
(NTC). \emph{A posteriori}, the best choice has proved to be the
latter~\menta{ttr}. Thermistors are electronic devices, hence conditioning
circuitry is necessary to drive them and acquire their data. The
requirement set by equation~(\ref{Lobo_eq.22}) is very demanding, indeed.
A test of the instrumentation is accordingly difficult, since best
laboratory thermal conditions are orders of magnitude less stable
than~(\ref{Lobo_eq.22}). We have devised an insulating system which
is capable to strongly screen ambient temperature fluctuations and
generate an environment in its interior where thermal stability is
10 times higher than~(\ref{Lobo_eq.21}), and more.

\begin{figure}[t]
\centering
\includegraphics[width=11cm]{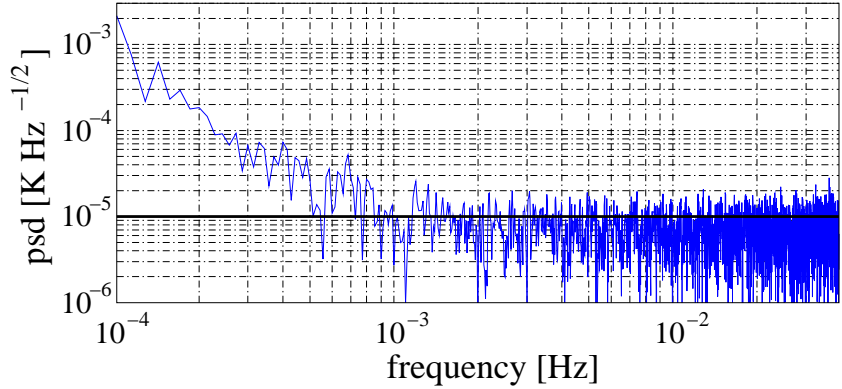}
\caption{Spectral density data of NTC temperature sensors. Note full
compliance with the requirement~(\protect\ref{Lobo_eq.22}) throughout
the measuring bandwidth.
\label{Lobo_fig.11}}
\end{figure}

Tests of the electronics and sensors are long duration. This is
understood in terms of the large time constants of the insulator, but
also because our measuring bandwidth (MBW) is around 1~mHz, hence periods
in the order of hours. Many tests have been conducted in the last two
years~\menta{nos1}, with different sensors and setups, which have been
used to constantly improve the performance of the temperature sensing
system. Thermistors consistently gave one order of magnitude less noise
than platinum resistors, and an example result is shown in
figure~\ref{Lobo_fig.11}.

\subsection{Calibration heaters}

As already explained, the key to using temperature sensors is to know
their calibration properties. This is accomplished by means of heaters
which generate suitable signals. The type of signals and the powers
required for each set of heaters varies. We separately summarise the
most relevant results.

\subsubsection{Inertial sensor heaters}

\begin{figure}[t]
\centering
\includegraphics[width=14cm]{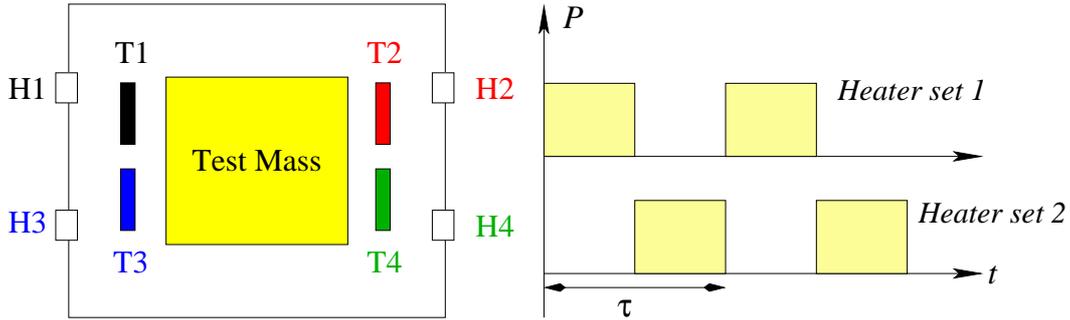}
\caption{Heaters for the Inertial Sensor Heads (left) and their activation
signals (right).}
\label{Lobo_fig.12}
\end{figure}

A schematic diagramme is shown in figure~\ref{Lobo_fig.12}, left: H1
through H4 are heaters, attached to the electrode housing. Colored
rectangles labeled T1 through T4 are the electrodes, and T1,\ldots,T4
their temperatures. To be noted that the heaters H1 and H3 form a
\emph{single logical heater}, which means they are driven at the same
time by identical signals. The same applies to heaters H2 and H4.

GRS, or Inertial Sensor heaters activation scheme is shown in
figure~\ref{Lobo_fig.12}, right. It is actually required that both the
period $\tau$ and the \emph{duty cycle} (ratio between the duration of
the ON and OFF signal periods) be tunable by software. This ensures that
the \ltp bandwidth is scanned, which we need to fully characterise the
system response. It is the Spanish DDS team's responsibility to define
the optimum power and activation scheme for the \ltp heaters. In the
case of the ISH heaters this had to be done on the basis of simulation
work, which is the only tool available at a time when the hardware is
still under construction. Simulations are done with a software toolkit
developed by \textsl{Carlo Gavazzi Space} (CGS) for the current
\ltp configuration.

\begin{figure}[b]
\centering
\includegraphics[width=11.3cm]{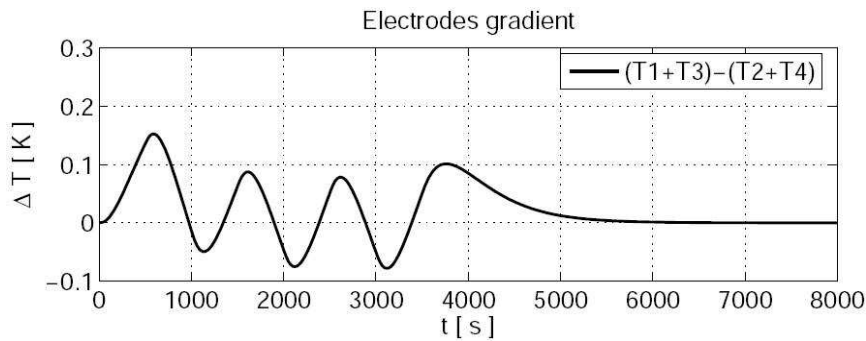}
\caption{Gradient response across the electrode housing for an alternate
signal of 500+500 seconds.}
\label{Lobo_fig.13}
\end{figure}

Figure~\ref{Lobo_fig.13} is an example of simulation results, corresponding
to $\tau$\,=\,10$^3$ seconds, duty cycle of 50\,\%~\menta{miquel}.
Imposing that signal-to-noise ratio be 50, the power of the heaters
can be calculated. In this case, the CGS model indicates that 5~mW
suffice. Similar methods applied to higher frequency signals show
that powers are higher, as indeed expected. At the band's higher end,
i.e., 30~mHz, 90~mW are needed. This is thus the requirement for
ISH heaters' powers.

\subsubsection{Optical Window heaters}

\begin{figure}[t]
\centering
\includegraphics[width=6.2cm]{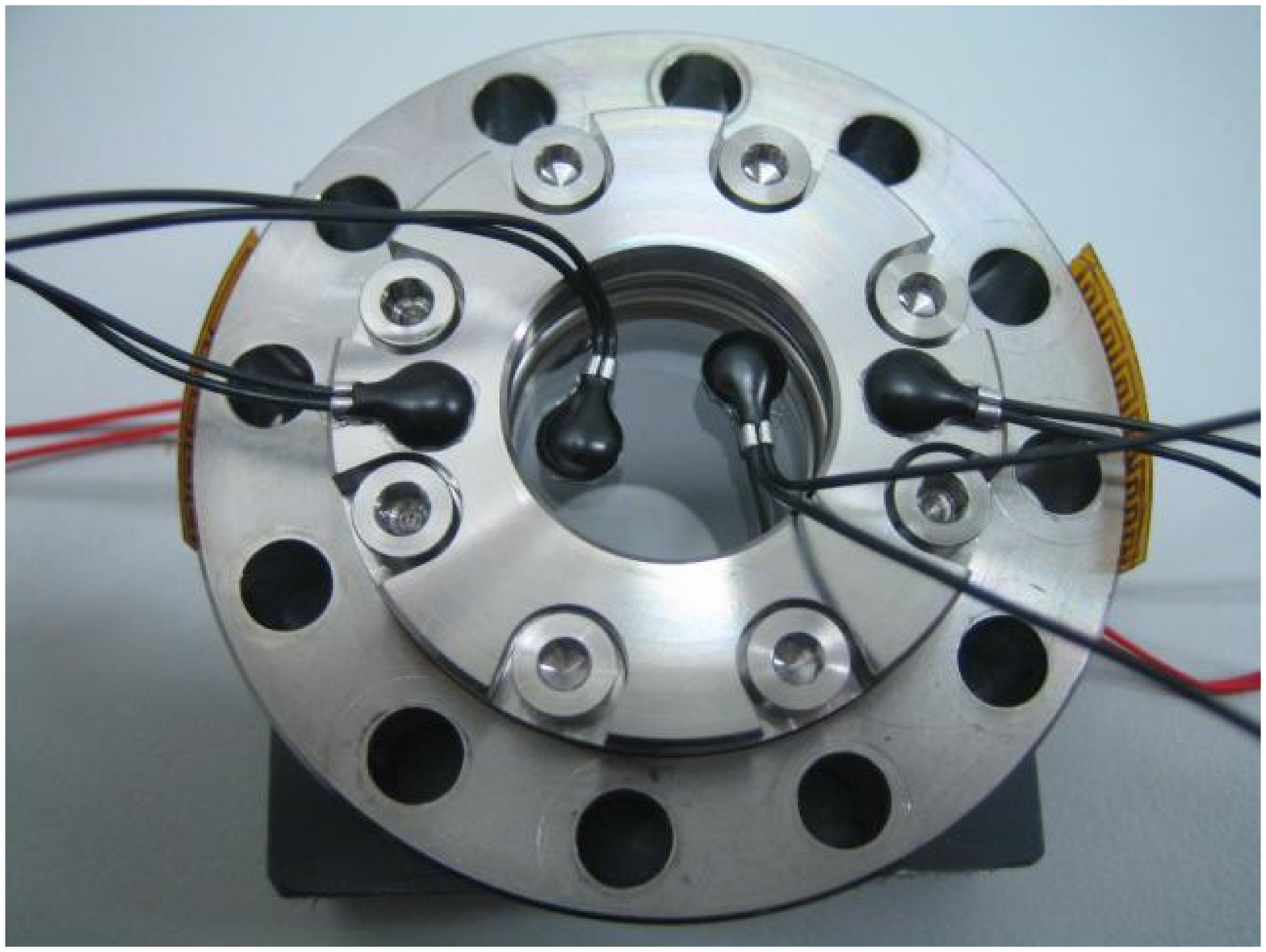}\qquad
\includegraphics[width=6.7cm]{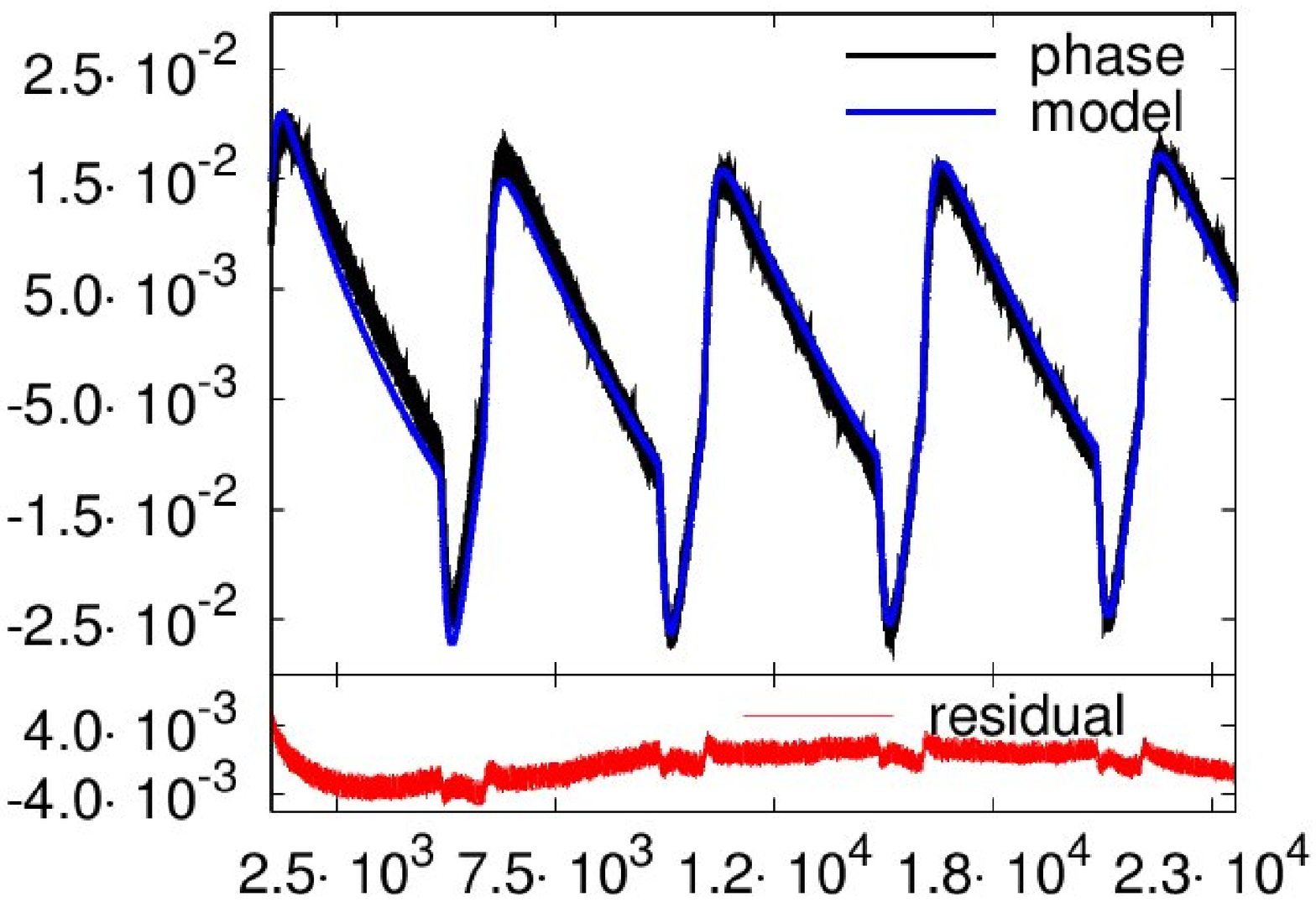}
\caption{The optical window, held in place on the titanium flange (left).
On the right, a plot comparing phase data with a theoretical model.}
\label{Lobo_fig.14}
\end{figure}

CGS's thermal model does not include the other important location for the
placement of heaters: the optical windows (OW). Experiment has therefore
been resorted to in order to properly characterise the OW response to
heat shocks. To this end, a collaboration between Barcelona and
Hannover was programmed to do an \emph{in situ} series of measurements.
The layout is seen in figure~\ref{Lobo_fig.14}, left: the temperature sensors
are four small black beads, two on the titanium and two on the glass.
Heaters are less visible, as they are orange-colored kapton tapes,
glued to the lateral surface of the titanium flange. Many data were
gathered, and offline analysis is quite advanced. A good example is
shown in figure~\ref{Lobo_fig.14}, right. The fit was done with a
form of auto-regressive process~\menta{migodd}:
\begin{equation}
 \phi(t) = \frac{a_1+a_2\,q^{-1}}{b_1+b_2\,q^{-1}}\;x(t)
 \label{Lobo_eq.23}
\end{equation}
where $x(t)$ is some linear combination of the temperatures read by the
sensors, and $q\/$ is the shift operator: $qx_n$\,$\equiv$\,$x_{n+1}$,
$q^{-1}x_n$\,$\equiv$\,$x_{n-1}$. $\phi(t)$ is the phase shift induced
by the thermal signals (pulsed shocks) on the light going through the
optical windows ---determined by interferometry.

\section{Magnetic diagnostics}

The \ltp test masses are two cubes 4.6 cm to the side, weighing 1.96~kg
each. They are made of an alloy of gold and platinum with
70\,\% Au + 30\,\%\ Pt, which is a compromise between weak magnetic
properties and mechanical resistance: magnetic-wise,
90\,\% Au + 10\,\% Pt would be better, but too high a ratio of gold
would make the cubic block vulnerable to mechanical deformation during
severe launch shocks.

To cast such an alloy is a process where ferromagnetic impurities can
enter the alloy structure, thus leaving a remnant magnetic moment in
the TM. Likewise, magnetic susceptibility will be present. State of
the art measurement techniques for these magnitudes is the driver for
a requirement on them, which is set to~\menta{tlsr}:
\begin{equation}
 |\chi| < 10^{-5}\ ,\quad
 |{\bf m}_0| < 10^{-8}\ {\rm Am}^2
 \label{eq.11}
\end{equation}

With these magnetic properties, magnetic cleanliness requirements must
be defined which ensure that no force and/or torque noise distorts the
position of the test masses. Let us see how these requirements come
about.

The magnetic force on a dipole of volume $V$, with roughly uniform
magnetic susceptibility $\chi$ and remnant magnetisation {\bf M}, is
given by the textbook formula
\begin{equation}
 {\bf F} =
 \left\langle\left[\left({\bf m}_0+
 \frac{\chi V}{\mu_0}\,{\bf B}\right)\cdot\nabla\right]\,{\bf B}\right\rangle
 \label{Lobo_eq.25}
\end{equation}
where $\langle\cdots\rangle$ indicates space average over the volume
of the test mass, and where {\bf m}$_0$\,=\,{\bf M}$V$.

The important feature to stress here is the \emph{quadratic} dependence
of the force on the magnetic induction field {\bf B}, which is due to
the susceptibility being different from zero, i.e., to the induced
magnetic moment $(\chi V/\mu_0)\,{\bf B}$.
The consequence of this is that magnetic force \emph{fluctuations} also
depend on DC values of the magnetic field. Requirements must accordingly
be defined for \emph{both} DC and fluctuating values of the magnetic field.
Various compromises among different noise contributions result in the
following table~\menta{tlsr}:

\begin{center}
\begin{tabular}{lcrl}
DC Magnetic field: & \quad & 10 & $\mu$T \\
DC Magnetic field gradient: & \quad & 5 & $\mu$T/m \\
Magnetic field fluctuations: & \quad & 650 & nT/$\sqrt{\rm Hz}$ \\
Magnetic field gradient fluctuations: & \quad & 250 & (nT/m)/$\sqrt{\rm Hz}$
\end{tabular}
\end{center}

The Industrial Architect must of course ensure that these requirements
of magnetic cleanliness are met. As regards diagnostics, as already
explained, the philosophy is different: we want to make magnetic
measurements to gain as precise knowledge as possible of the values
of all relevant magnetic magnitudes. We come to this now.

\subsection{Magnetic field measurement}

\begin{figure}[t]
\centering
\includegraphics[width=10cm]{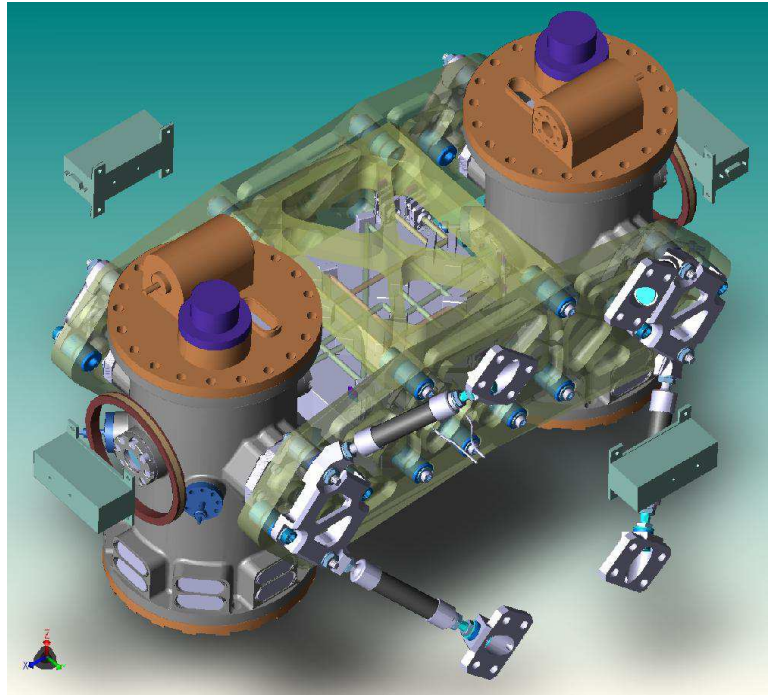}
\caption{The external look of the \ltp. Magnetometers and coils are drawn
levitating to avoid obscuring the figure with the mechanical interfaces.}
\label{Lobo_fig.15}
\end{figure}

Magnetometers are often instruments based on high magnetic permeability
cores which are submitted to saturation hysteresis cycles. These magnetic
cores can severely challenge the magnetic cleanliness inside the \ltp,
and should therefore be kept safely away from the test masses. The most
sensitive devices usable with the \ltp are \emph{fluxgate} magnetometers,
which also have a magnetic core\footnote{
There are more sensitive devices, for example SQUIDs, but these require
cryogenic conditions, hence cannot be possibly flown with the \ltp due
to unacceptable pump mechanical noise.}. In addition, very sensitive
fluxgates are usually heavy. All in all, we are forced to use only a
few sensors, and far from the test masses. The consolidated requirement
is that \emph{four} be used, in a configuration which we see in
figure~\ref{Lobo_fig.15}. These constraints result in problems to create
an accurate field map in the region where the test masses are, since an
extrapolation of the magnetometers' readouts does not produce sufficient
information, and (debatable) criteria must be applied on how to best
make use of the available data.

Studies developed by us have provided an approach to address the problem
of magnetic field extrapolation~\menta{atip}. It is based on the datum
that the sources of magnetic disturbance in the entire spacecraft are
known, i.e., we know \emph{where} they are, and their nominal
\emph{dipole moments}. What is of course unknown is their
\emph{fluctuations}. If all the magnetic dipole moments were actually
known then one could reconstruct the magnetic field map by simple linear
superposition of the fields of each dipole ---but this of course does
not happen in practice.

There are 51 identified sources, most of them PCBs (printed circuit boards),
which means their magnetic dipoles are perpendicular to each board's plane.
This considerably reduces the number of unknowns, since knowing the 51
\emph{moduli} of the dipole moment vectors suffices to completely identify
them. This simplifies the problem, but we are still far from done: to
determine 51 magnitudes from 12 experimental data is an under-determined,
or \emph{degenerate} problem: a multiply infinite set of solutions exists
for it. The question is thus: can one solution be selected out of so many?

We have implemented a sort of least squares method for such selection,
where random sets of solutions are iteratively compared, and a new selection
is made in each step. The method is seen to converge, though final values
have errors within up to 50\,\% of the real result. The analysis process is
of course to be performed offline, and we are still working on improvements.

\subsection{Magnetic coils}

The purpose of the magnetic coils is to generate strong signals which
cause forces to act on the TMs. Magnetic forces happen in non-uniform
magnetic fields, so a single coil per TM has been selected for in-flight
operations ---this is the simplest scheme, see Figure~\ref{Lobo_fig.15}.

Contrary to the situation with heaters, magnetic coils have an effect
on the TMs which can be determined with high precision thanks to the
well known laws magnetic induction by loop currents. This being the
case, coils will be used to determine the magnetic properties of the
TMs, which can later be used to calculate the magnetic noise contribution
based on the magnetometers' readings and the magnetic field evaluation
described in the previous section.

Because of the quadratic dependence of the magnetic force on the
magnetic field, the result of applying an oscillating current of well
known frequency, $\omega$, to the coils is that the force acting on the
TMs has two Fourier components: one at the same frequency $\omega$ and
one at double that frequency, 2$\omega$. It is readily seen that
\begin{eqnarray}
 \bma{F}_\omega & = & (\bma{M\cdot\nabla})\bma{B}_{\rm app} +
 \frac{\chi V}{\mu_0}\,\left[(\bma{B}_{\rm ext}\bma{\cdot\nabla})
 \bma{B}_{\rm app} + (\bma{B}_{\rm app}\bma{\cdot\nabla})\bma{B}_{\rm ext}
 \right] \label{Lobo_eq.26a} \\
 \bma{F}_{2\omega} & = & \frac{\chi V}{\mu_0}\,
 (\bma{B}_{\rm app}\bma{\cdot\nabla})\bma{B}_{\rm app}
 \label{Lobo_eq.26b}
\end{eqnarray}
where
\begin{equation}
 \bma{B}_{\rm app} = \bma{B}_0\,e^{i\,\omega t}
 \label{Lobo_eq.27}
\end{equation}
is the applied field of the coils, and $\bma{B}_{\rm ext}$ is the otherwise
prevailing field in the TMs.

Equation~(\ref{Lobo_eq.26b}) shows that measurement of the system response at
the double frequency, 2$\omega$, suffices to determine the susceptibility
of the test mass. To measure the remnant magnetisation $\bma{M}$ is more
complicated, as a \emph{vector} measurement at frequency $\omega$ needs
to be done. Difficulties are however expected to be mitigated by previous
on-ground measurements of these quantities, which should not be too
affected by launch trauma.

The coils will be aligned with the line joining the test masses. They
are 113~mm in diameter, 9~mm thick and are 81.5~mm away from the TMs.
They have 2400 turns and will be fed by a current of a few milli-Amperes.

\section{Radiation Monitor}

Cosmic rays and certain solar events contain ionising particles which
will hit the \textsl{LTP} in flight, thus causing spurious signals in
the GRS. These particles are mostly protons, with 10\,\% or less of
He nuclei, and a minor component of heavier galactic nuclei and solar
ions. Charging rates and the properties of noise caused by charging vary
depending on whether the particle flux comes from Galactic Cosmic Rays
(\textsl{GCR}) or is augmented by Solar Energetic Particles (\textsl{SEP}).
The reason is that the two types of radiation present different energy
spectra, which result in different TM charging efficiencies. Temporal
\emph{fluctuations} of the \textsl{GCR} flux and \textsl{SEP} fake GW
signals, and therefore a particle counter is necessary to provide
correlations between the flux of energetic particles and the
instantaneous charging rates observed in the test masses. The device
must be able to determine the \emph{energy spectra} of the detected
particles, as this is the natural tool to tell \textsl{SEP} events from
\textsl{GCR}s.

Not all charged particles hitting the satellite structure will make it to the
test masses, as that structure itself has a certain \emph{stopping power}.
The particle counter must only be triggered by those particles having
enough energy to reach the TMs, hence it must be properly \emph{shielded}.
Simulation work indicates that only ions with energies larger than
$\sim$100\,MeV should be counted~\menta{iclsimu}. The particle counter
together with the above added capabilities is known as Radiation Monitor (RM).

\begin{figure}[t]
\centering
\includegraphics[width=9cm]{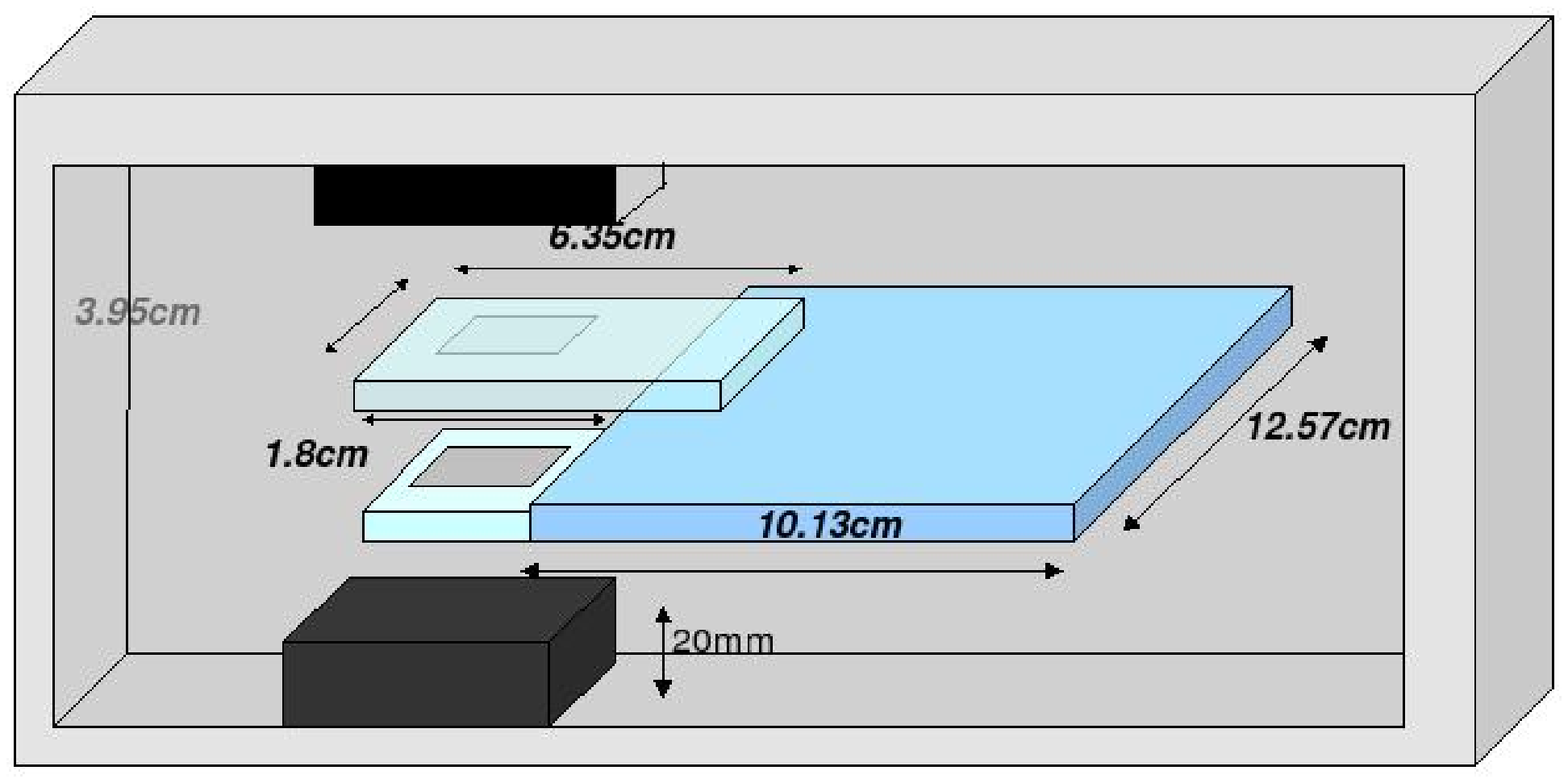}\qquad
\includegraphics[width=4.3cm]{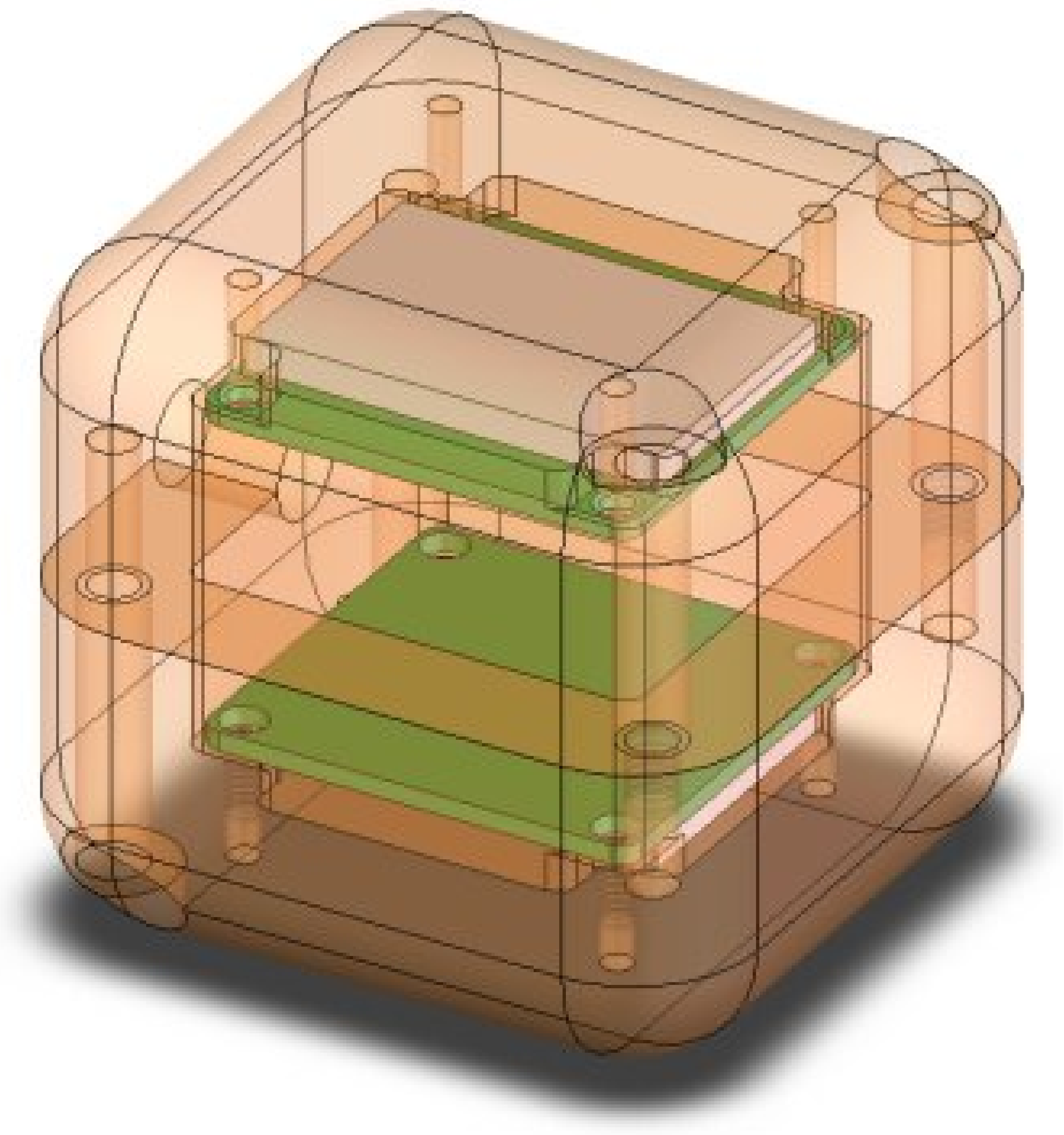}\qquad
\caption{Radiation Monitor concept: two PIN diodes in telescopic
configuration (left). The PIN diodes are shielded with a thick metallic
layer (right).
\label{Lobo_fig.16}}
\end{figure}

The idea of the RM was first developed at Imperial College~\menta{ha}.
It is based on two PIN diodes, each of which can count individual particle
hits. Photons are absorbed, while protons and heavier nuclei can exit
through the opposite side of the diode. If placed in a telescopic
configuration (see figure~\ref{Lobo_fig.16}, left), the system can register
coincidences and calculate spectra of the charged incident particles based
on energy deposition in the PIN diodes. There is some degeneracy of course,
due to the monitor's acceptance angle.

The actual implementation in hardware of these ideas has been carried
through by a team at IFAE~\menta{cesar}.

\begin{figure}[t]
\centering
\includegraphics[width=12cm]{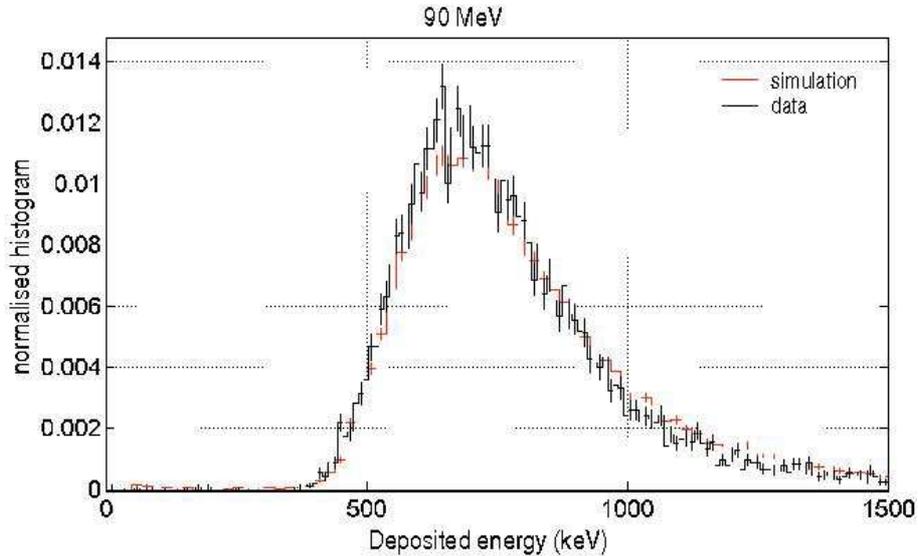}
\caption{Comparison of deposited energy spectra from simulation (red) and
test beam data (black) for 90~MeV protons.
\label{Lobo_fig.17}}
\end{figure}

Contrary to the previous diagnostics, the RM does not require in-flight
calibration. This is done on ground by submitting it to laboratory
proton beam irradiation. An experiment was prepared in the Paul Scherrer
Institute (PSI, Switzerland) in November 2005 to check that the shield was
working as expected ---previous tests of the electronics had of course
been done earlier. An example result is shown in figure~\ref{Lobo_fig.17},
which shows very satisfactory confirmation that simulation predictions were
correct. To obtain such results was however not immediately straightforward,
as various simulation and hardware parameters had to be properly tuned
to give proper account of a number of non-obvious experimental
facts~\menta{wass}.

An unexpected potential complication emerged out of the November 2005
proton beam run: evidence was found that, under heavy proton flux
exposure (up to $\sim$10$^{10}$ protons), copper gets activated.
A new PSI run was set up in November 2006. Final conclusions are still
pending at the time of writing. We do not expect major RM shield changes
to be required, but the results of the analysis will be necessary for a
thorough understanding of the RM~readout.

\section{Conclusion}

The history of GW detection science is almost 50 years old. The long
striven for detection has not been successful yet, despite the life
long efforts of already two generations of researchers world wide.
The reward to these people, and their funding bodies, has surely been
the confirmation of truly impressive progress in instrument sensitivity.
Laboratory detectors have limitations bound to their being on the Earth
surface, and this constrains their viability to GW signals with spectra
in the kHz band. \lisa will be free from such limitations by going to
outer space, thence making observations at much lower frequencies, in
the milli-Hz band. Even if, as we expect, Earth detectors see GW signals
well before \lisa flies, this joint \esa-\nasa mission will be a unique
GW observatory, specific to the low frequency spectrum.

\lisa is of course a complicated mission. Even though its scientific case
is almost universally acknowledged, the difficulties of its implementation
initially caused numerous caveats about its real viability. Nevertheless
enthusiasm for \lisa has mounted internationally in the last five years
or so: the community is now well organised into working groups, and the
number of papers on \lisa has soared. But no less engineering activities
are thriving, too, as the precursor \lisa \textsl{PathFinder} approaches
its definitive implementation stages. \lpf successfully completed its
Preliminary Design Review (PDR), and is now going through Critical Design
Review (CDR), the last stage before full speed manufacture begins. Current
plans schedule \lpf launch for late 2009, and indications are positive
about feasibility. The challenge will then be the analysis of the mission
results, which will be crucial for \lisa. We have good hopes that success
will come, and this paper intends to contribute evidence that such hopes
are based on reason, not on wishful thinking.

\ack

We thank the Spanish \emph{Ministerio de Educaci\'on y Ciencia} for
support, contract {\sf ESP2004-01647.}

\end{document}